\documentclass[journal]{IEEEtran}
\ifCLASSINFOpdf
  \usepackage[pdftex]{graphicx}
   \graphicspath{{../figures/}{../jpeg/}}
\else
\fi

\usepackage[cmex10]{amsmath}
\usepackage{amssymb}
\usepackage{bm}

\usepackage{algpseudocode}
\usepackage{algorithm}
\usepackage{array}
\usepackage{url}
\usepackage[noadjust]{cite}

\newtheorem{proposition}{Proposition}
\newtheorem{remark}{Remark}

\usepackage{fancyhdr}
\usepackage{amsfonts}
\usepackage{xfrac}
\usepackage{color}
\usepackage{pgfplots}
\usepackage{pgfplots}
\pgfplotsset{compat=1.17} 
\pgfplotsset{plot coordinates/math parser=false} 
\pgfplotsset{filter discard warning=false} 
\pgfplotsset{table/col sep=space} 
\usepackage{enumerate}

\usepackage{morewrites}
\usepackage{morefloats}
\usepackage{expl3}
\usepackage{shellesc} 

\usepackage{cite}
\usepackage{mathrsfs}
\usepackage{subcaption}
\usepackage{stfloats}
\usepackage{makecell}
\usepackage{comment}
\usepackage{afterpage}
\usepackage{tikz}
\usepackage{mathdots}
\usepackage{yhmath}
\usepackage{cancel}
\usepackage{color}
\usepackage{siunitx}
\usepackage{array}
\usepackage{multirow}
\usepackage{textcomp}
\usepackage{gensymb}
\usepackage{tabularx}
\usepackage{booktabs}
\usepackage{graphicx}
\usepackage{xcolor}
\usepackage{mathtools}
\newtagform{normalsize}[\normalsize]{\normalsize(}{\normalsize)}

\allowdisplaybreaks
\usetikzlibrary{decorations.markings,decorations.pathmorphing}

\begin{document}
    \title{How Many Pinching Antennas Are Enough?}
  \author{Dimitrios~Tyrovolas,~\IEEEmembership{Member,~IEEE,} 
      Sotiris A. Tegos,~\IEEEmembership{Senior Member,~IEEE,} Yue Xiao,~\IEEEmembership{Member,~IEEE,}  \\ Panagiotis D. Diamantoulakis,~\IEEEmembership{Senior Member,~IEEE,} Sotiris Ioannidis, Christos K.~Liaskos,~\IEEEmembership{Member,~IEEE,} \\ George K.~Karagiannidis,~\IEEEmembership{Fellow,~IEEE} and~Stylianos D. Asimonis,~\IEEEmembership{Senior Member,~IEEE}
 
\thanks{D. Tyrovolas is with the Department of Electrical and Computer Engineering, University of Patras, 26504 Patras, Greece and with Dienekes SI IKE Heraklion, Crete, Greece (dtyrovolas@upatras.gr).}
\thanks{S. A. Tegos, P. D. Diamantoulakis, and G. K. Karagiannidis are with the Department of Electrical and Computer Engineering, Aristotle University of Thessaloniki, 54124 Thessaloniki, Greece(tegosoti@auth.gr, padiaman@auth.gr, geokarag@auth.gr).}
\thanks{Yue Xiao is with the School of Information Science and Technology, Southwest Jiaotong University, 610031, Chengdu, China  (alice\_xiaoyue@hotmail.com).}
\thanks{S. Ioannidis is with the Dept. of Electrical and Computer Engineering, Technical University of Chania, Chania, Greece and with Dienekes SI IKE Heraklion, Crete, Greece (sotiris@ece.tuc.gr).}
\thanks{C. K. Liaskos is with the Computer Science Engineering Department, University of Ioannina, Ioannina, and Foundation for Research and Technology Hellas (FORTH), Greece (cliaskos@ics.forth.gr).} 
\thanks{S. D. Asimonis is with the Department of Electrical and Computer Engineering, University of Patras, 26504 Patras, Greece (s.asimonis@upatras.gr).}
\thanks{This work has been funded by the European Union’s Horizon 2020 research and innovation programs under grant agreement No 101139194 6G Trans-Continental Edge Learning (6G-XCEL). The work of C. K. Liaskos has received funding from the SNS-JU under the European Union’s Horizon Europe research and innovation programme, in the frame of CYBERSECDOME (No 101120779).}
}
\maketitle

\begin{abstract} 
Programmable wireless environments (PWEs) have emerged as a key paradigm for next-generation communication networks, aiming to transform wireless propagation from an uncontrollable phenomenon into a reconfigurable process that can adapt to diverse service requirements. In this framework, pinching-antenna systems (PASs) have recently been proposed as a promising enabling technology, as they allow the radiation location and effective propagation distance to be adjusted by selectively exciting radiating points along a dielectric waveguide. However, most existing studies on PASs rely on the idealized assumption that pinching-antenna (PA) positions can be continuously adjusted along the waveguide, while realistically  only a finite set of pinching locations is available. Motivated by this, this paper analyzes the performance of two-state PASs, where the PA positions are fixed and only their activation state can be controlled. By explicitly accounting for the spatial discreteness of the available pinching points, closed-form analytical expressions for the outage probability and the ergodic achievable data rate are derived. In addition, we introduce the pinching discretization efficiency to quantify the performance gap between discrete and continuous pinching configurations, enabling a direct assessment of the number of PAs required to approximate the ideal continuous case. Finally, numerical results validate the analytical framework and show that near-continuous performance can be achieved with a limited number of PAs, offering useful insights for the design and deployment of PASs in PWEs.
\end{abstract}

\maketitle

\section{Introduction}

In next-generation wireless networks, both industry and academia are converging on the vision of intelligent wireless systems capable of supporting heterogeneous applications such as immersive extended reality, advanced healthcare, and the metaverse \cite{Holographic, healthcare, 6GMetaverse}. These emerging use cases impose stringent requirements on reliability, latency, and data rate, while simultaneously demanding that the wireless environment dynamically adapt to multiple users with diverse objectives and quality of service requirements. At the same time, the use of high frequency bands, although offering large bandwidths, intensifies fundamental challenges such as severe path loss, sensitivity to blockage, and rapid channel variations, all of which can significantly degrade communication performance \cite{6GNetwork, Donal2024}. To address these challenges, programmable wireless environments (PWEs) have emerged as a paradigm in which electromagnetic wave propagation is no longer treated as an uncontrollable phenomenon, but rather as a software defined process that can be shaped according to system objectives \cite{PWELiaskos}. In this context, key parameters that traditionally degrade quality of service (QoS), including path loss and fading, are envisioned to become controllable and reconfigurable design variables, which motivates the need to identify and develop reconfigurable technologies capable of controlling large-scale and small-scale propagation characteristics.

Among the technologies envisioned to support PWEs, pinching antennas (PAs) have recently emerged as a particularly promising approach to transform path loss into a reconfigurable parameter \cite{book,DOCOMO}. In more detail, by utilizing dielectric waveguides to guide electromagnetic waves at high frequencies and dynamically creating radiating points through localized dielectric perturbations, PAs enable flexible adjustment of antenna placement along the waveguide, thus allowing the effective propagation distance and radiation footprint to be adapted to user location and environmental geometry \cite{DOCOMO,pinchingMAG,YuanweiMag2025}. This unique operating principle allows PAs to establish adjustable line of sight links even in complex or obstructed environments without requiring additional active hardware, making them a practical and efficient solution for dynamically shaping wireless propagation \cite{Ding2024TCOM,TegosPinching}. Therefore, it becomes essential to examine the performance capabilities of PAs under practical deployment conditions in order to establish a realistic understanding of their achievable benefits.

\subsection{State-of-the-Art}

An increasing amount of literature has focused on understanding the fundamental operation and performance limits of pinching-antenna systems (PASs) across a variety of communication scenarios \cite{pinchingMAG, Ding2024TCOM}. In this direction, early studies have primarily concentrated on modeling the electromagnetic behavior of dielectric waveguides with dynamically activated radiating points, establishing tractable signal and channel models that capture the interaction between waveguide propagation, radiation phenomena, and free-space transmission \cite{PASS,Modeling2025}. These works provide key insights into how pinching locations shape the effective channel response and demonstrate that PASs can achieve significant array and beamforming gains. Following these foundational modeling efforts, several works have focused on analytically characterizing the performance of PASs and identifying the key parameters that define their behavior. Specifically, the authors in \cite{TyrovolasPASS2025} present a comprehensive performance analysis of PASs, deriving closed-form expressions for outage probability and achievable rate that explicitly reveal the impact of waveguide length and dielectric losses, illustrating that performance gains are closely tied to the balance between waveguide attenuation and propagation distance. Moreover, the authors of \cite{PerformancePASSSWIPT2025} analyze PA-enabled systems under simultaneous information and energy transfer and emphasize on how the length and placement of the dielectric waveguide influence system performance through large-scale propagation effects. Furthermore, the authors of \cite{VasilisPASS} investigate wireless-powered PAS networks and show that the spatial adaptability of PAs can be exploited to mitigate the double near-far problem, thereby improving both fairness and energy efficiency. Finally, in \cite{zhou2025}, a general modeling framework for beamforming design in downlink PASs was introduced for users equipped with multiple receive antennas. 

As interest in PASs has expanded, research has increasingly examined how pinching-based reconfigurability affects system behavior in multi-user communication and resource allocation settings. In more detail, the integration of PASs with non-orthogonal multiple access in \cite{NOMAPASS, NOMAPinch, xu2025qosawarenomadesigndownlink} shows that antenna activation along the waveguide can be jointly designed with power allocation to manage inter-user interference, revealing how the reconfigurability of PASs can influence throughput in multi-user scenarios. This insight is further developed through the study of orthogonal frequency-division multiple access and rate-splitting multiple access in \cite{ThrassosPASS} and \cite{APOSTOLOSRSMA}, respectively, where adapting PA positions enables PASs to support diverse access mechanisms while preserving transmission efficiency. Beyond multiple access, the impact of PA adaptability is examined in the context of positioning \cite{PositioningPASS2025} and integrated sensing and communication in \cite{BozanisPASS, ISACPASSYuanwei2025}, where the PA positions jointly shape sensing accuracy and communication performance. Related insights into PAS characteristics are provided by studies on index modulation \cite{IndexModulationPASS2025} and physical layer security \cite{PIGIPASS}, which illustrate how the spatial degrees of freedom introduced by PAs influence signal distinguishability and information leakage. Finally, works on segmented waveguide architectures and waveguide-level transmission strategies, such as SWANs and waveguide-division or waveguide-multiplexing designs \cite{SWAN2025,WDMAPASS2025}, demonstrate how the PAS architecture itself can be further exploited to support scalable multi-user transmission through waveguide-level multiplexing. Overall, these studies emphasize that the PAS performance is fundamentally influenced by the interaction between waveguide propagation, spatial reconfiguration through pinching, and the surrounding communication environment.

\subsection{Motivation \& Contribution}
Despite substantial progress in understanding and exploiting the reconfigurability of PASs, most existing works rely on the idealized assumption that PAs can be continuously adjusted along dielectric waveguides. In practical implementations, however, only a finite number of pinching locations can be realized, which fundamentally alters the structure of optimization and performance analysis. In this direction, recent studies have begun to examine two-state PASs, where the positions of the PAs are fixed along the waveguide and only their activation state can be controlled. In particular, the authors of \cite{DISCRETEMECPASS2025} examined a two-state PAS in a wireless-powered mobile edge computing scenario to compare multiple-access strategies, while the authors of \cite{KAIKITPASS2025} investigated dynamic PA placement for multicast downlink transmission and demonstrated that discrete PA positions can still yield significant performance gains. More recently, learning-based approaches have been proposed in \cite{ODYSSEAS2025} to optimize two-state PASs, illustrating the potential of data-driven methods to handle the combinatorial nature of discrete antenna activation. However, while these works confirm the practical relevance of two-state PASs, they primarily rely on numerical optimization or learning-based techniques and therefore offer limited analytical insight into how discreteness, waveguide attenuation, and user location jointly shape system performance. To the best of the authors' knowledge, no closed-form mathematical framework has yet been developed that explicitly quantifies the impact of discrete pinching locations on key performance metrics of PASs to provide direct design insights.

In this work, we analyze the performance of two-state PAS, where the positions of the PAs are fixed along the dielectric waveguide and only their activation state can be controlled. By explicitly accounting for the spatial discreteness of the available pinching points, we derive closed-form analytical expressions for the ergodic data rate, which provide an exact characterization of the achievable rate performance under this practical deployment model. In addition, we introduce the notion of \emph{pinching discretization efficiency} (PDE) in order to quantify the performance gap between discrete and continuous pinching configurations, allowing a direct assessment of how the number of available PAs affects the achievable performance. This analysis establishes a clear connection between the number of PAs, the system geometry, and the resulting rate performance, offering useful insights for the design and deployment of PASs in PWEs.

\subsection{Structure}
The remainder of the paper is organized as follows. The system model is described in Section II, while the performance analysis is given in Section III. Finally, our results are presented in Section IV, and Section V concludes the paper.

\section{System model}

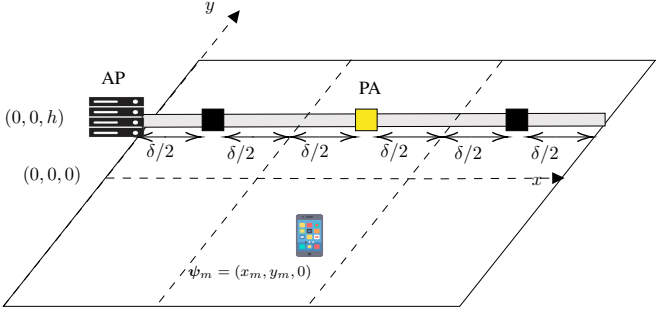
\begin{figure}
 \centering
    \resizebox{\columnwidth}{!}{
\begin{tikzpicture}[x=0.75pt,y=0.75pt,yscale=-1,xscale=1]
\draw   (238.67,50) -- (566,50) -- (425.71,226.6) -- (98.38,226.6) -- cycle ;
\draw  [dash pattern={on 4.5pt off 4.5pt}]  (98.38,226.6) -- (266.13,16.35) ;
\draw [shift={(268,14)}, rotate = 128.58] [fill={rgb, 255:red, 0; green, 0; blue, 0 }  ][line width=0.08]  [draw opacity=0] (8.93,-4.29) -- (0,0) -- (8.93,4.29) -- cycle    ;
\draw  [fill={rgb, 255:red, 235; green, 232; blue, 233 }  ,fill opacity=1 ] (530,88) -- (197.97,88.87) -- (198,98) -- (530.03,97.13) -- cycle ;
\draw  [fill={rgb, 255:red, 248; green, 231; blue, 28 }  ,fill opacity=1 ] (351,85) -- (366.5,85) -- (366.5,100.5) -- (351,100.5) -- cycle ;
\draw (317.75,176) node  {\includegraphics[width=28.13pt,height=22.5pt]{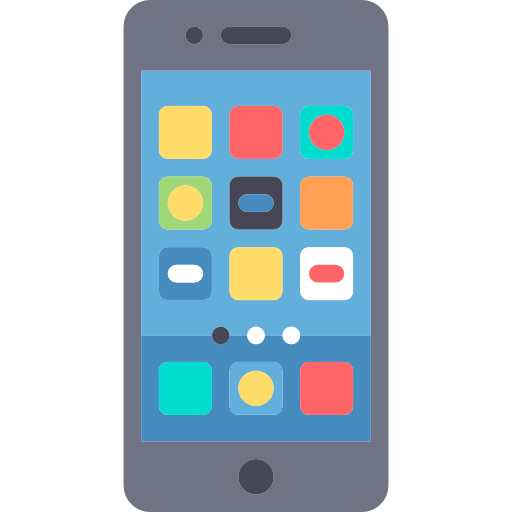}};
\draw  [dash pattern={on 4.5pt off 4.5pt}]  (172,134.5) -- (411.78,135.07) -- (498,134.52) ;
\draw [shift={(501,134.5)}, rotate = 179.63] [fill={rgb, 255:red, 0; green, 0; blue, 0 }  ][line width=0.08]  [draw opacity=0] (8.93,-4.29) -- (0,0) -- (8.93,4.29) -- cycle    ;
\draw    (215,105) -- (196.5,105) ;
\draw [shift={(194.5,105)}, rotate = 360] [color={rgb, 255:red, 0; green, 0; blue, 0 }  ][line width=0.75]    (10.93,-3.29) .. controls (6.95,-1.4) and (3.31,-0.3) .. (0,0) .. controls (3.31,0.3) and (6.95,1.4) .. (10.93,3.29)   ;
\draw    (214,105) -- (238,105) ;
\draw [shift={(240,105)}, rotate = 180] [color={rgb, 255:red, 0; green, 0; blue, 0 }  ][line width=0.75]    (10.93,-3.29) .. controls (6.95,-1.4) and (3.31,-0.3) .. (0,0) .. controls (3.31,0.3) and (6.95,1.4) .. (10.93,3.29)   ;
\draw  [draw opacity=0] (145.25,147) .. controls (145.25,144.65) and (147.15,142.75) .. (149.5,142.75) .. controls (151.85,142.75) and (153.75,144.65) .. (153.75,147) .. controls (153.75,149.35) and (151.85,151.25) .. (149.5,151.25) .. controls (147.15,151.25) and (145.25,149.35) .. (145.25,147) -- cycle ;
\draw (179.5,90.75) node  {\includegraphics[width=66.75pt,height=47.63pt]{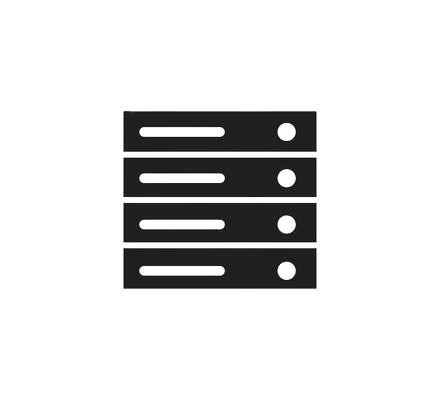}};
\draw  [fill={rgb, 255:red, 0; green, 0; blue, 0 }  ,fill opacity=1 ] (241,85) -- (256.5,85) -- (256.5,100.5) -- (241,100.5) -- cycle ;
\draw    (278,105) -- (259.5,105) ;
\draw [shift={(257.5,105)}, rotate = 360] [color={rgb, 255:red, 0; green, 0; blue, 0 }  ][line width=0.75]    (10.93,-3.29) .. controls (6.95,-1.4) and (3.31,-0.3) .. (0,0) .. controls (3.31,0.3) and (6.95,1.4) .. (10.93,3.29)   ;
\draw    (277,105) -- (301,105) ;
\draw [shift={(303,105)}, rotate = 180] [color={rgb, 255:red, 0; green, 0; blue, 0 }  ][line width=0.75]    (10.93,-3.29) .. controls (6.95,-1.4) and (3.31,-0.3) .. (0,0) .. controls (3.31,0.3) and (6.95,1.4) .. (10.93,3.29)   ;
\draw    (325,105) -- (306.5,105) ;
\draw [shift={(304.5,105)}, rotate = 360] [color={rgb, 255:red, 0; green, 0; blue, 0 }  ][line width=0.75]    (10.93,-3.29) .. controls (6.95,-1.4) and (3.31,-0.3) .. (0,0) .. controls (3.31,0.3) and (6.95,1.4) .. (10.93,3.29)   ;
\draw    (325,105) -- (349,105) ;
\draw [shift={(351,105)}, rotate = 180] [color={rgb, 255:red, 0; green, 0; blue, 0 }  ][line width=0.75]    (10.93,-3.29) .. controls (6.95,-1.4) and (3.31,-0.3) .. (0,0) .. controls (3.31,0.3) and (6.95,1.4) .. (10.93,3.29)   ;
\draw    (387,105) -- (368.5,105) ;
\draw [shift={(366.5,105)}, rotate = 360] [color={rgb, 255:red, 0; green, 0; blue, 0 }  ][line width=0.75]    (10.93,-3.29) .. controls (6.95,-1.4) and (3.31,-0.3) .. (0,0) .. controls (3.31,0.3) and (6.95,1.4) .. (10.93,3.29)   ;
\draw    (387,105) -- (411,105) ;
\draw [shift={(413,105)}, rotate = 180] [color={rgb, 255:red, 0; green, 0; blue, 0 }  ][line width=0.75]    (10.93,-3.29) .. controls (6.95,-1.4) and (3.31,-0.3) .. (0,0) .. controls (3.31,0.3) and (6.95,1.4) .. (10.93,3.29)   ;
\draw    (434,105) -- (415.5,105) ;
\draw [shift={(413.5,105)}, rotate = 360] [color={rgb, 255:red, 0; green, 0; blue, 0 }  ][line width=0.75]    (10.93,-3.29) .. controls (6.95,-1.4) and (3.31,-0.3) .. (0,0) .. controls (3.31,0.3) and (6.95,1.4) .. (10.93,3.29)   ;
\draw    (434,105) -- (458,105) ;
\draw [shift={(460,105)}, rotate = 180] [color={rgb, 255:red, 0; green, 0; blue, 0 }  ][line width=0.75]    (10.93,-3.29) .. controls (6.95,-1.4) and (3.31,-0.3) .. (0,0) .. controls (3.31,0.3) and (6.95,1.4) .. (10.93,3.29)   ;
\draw  [fill={rgb, 255:red, 0; green, 0; blue, 0 }  ,fill opacity=1 ] (459,85) -- (474.5,85) -- (474.5,100.5) -- (459,100.5) -- cycle ;
\draw    (495,105) -- (476.5,105) ;
\draw [shift={(474.5,105)}, rotate = 360] [color={rgb, 255:red, 0; green, 0; blue, 0 }  ][line width=0.75]    (10.93,-3.29) .. controls (6.95,-1.4) and (3.31,-0.3) .. (0,0) .. controls (3.31,0.3) and (6.95,1.4) .. (10.93,3.29)   ;
\draw    (495,105) -- (519,105) ;
\draw [shift={(521,105)}, rotate = 180] [color={rgb, 255:red, 0; green, 0; blue, 0 }  ][line width=0.75]    (10.93,-3.29) .. controls (6.95,-1.4) and (3.31,-0.3) .. (0,0) .. controls (3.31,0.3) and (6.95,1.4) .. (10.93,3.29)   ;
\draw  [dash pattern={on 4.5pt off 4.5pt}]  (348,49.5) -- (209.38,224.6) ;
\draw  [dash pattern={on 4.5pt off 4.5pt}]  (456,50.5) -- (317.38,225.6) ;

\draw (111,124.99) node [anchor=north west][inner sep=0.75pt]    {$( 0,0,0)$};
\draw (476.2,132.17) node [anchor=north west][inner sep=0.75pt]    {$x$};
\draw (242,6.72) node [anchor=north west][inner sep=0.75pt]    {$y$};
\draw (352.4,63.89) node [anchor=north west][inner sep=0.75pt]   [align=left] {PA};
\draw (229.7,195.89) node [anchor=north west][inner sep=0.75pt]  [font=\footnotesize]  {$\boldsymbol{\psi }_{m} =\left( x_{m} ,y_{m} ,0\right)$};
\draw (167.5,57) node [anchor=north west][inner sep=0.75pt]   [align=left] {AP};
\draw (98,83.99) node [anchor=north west][inner sep=0.75pt]    {$( 0,0,h)$};
\draw (199.75,108.4) node [anchor=north west][inner sep=0.75pt]    {$\delta /2$};
\draw (257.75,109.4) node [anchor=north west][inner sep=0.75pt]    {$\delta /2$};
\draw (310.75,108.4) node [anchor=north west][inner sep=0.75pt]    {$\delta /2$};
\draw (367.75,108.4) node [anchor=north west][inner sep=0.75pt]    {$\delta /2$};
\draw (420.75,109.4) node [anchor=north west][inner sep=0.75pt]    {$\delta /2$};
\draw (475.75,109.4) node [anchor=north west][inner sep=0.75pt]    {$\delta /2$};
\end{tikzpicture}
}
    \caption{Overview of two-state PAS.}
    \label{fig:system_model}
\end{figure}

We consider the downlink communication scenario depicted in Fig.~\ref{fig:system_model}, where an access point (AP) communicates with a single-antenna user located randomly within a rectangular area in the $x$-$y$ plane with dimensions $D_x$ and $D_y$. The user position is denoted by $\boldsymbol{\psi_m} = (x_m, y_m, 0)$, where $x_m$ is uniformly distributed over $[0, D_x]$ and $y_m$ is uniformly distributed over $\left[-D_y/2, D_y/2\right]$. To ensure reliable communication, the AP employs a dielectric waveguide which allows electromagnetic radiation from selected points along its length through a controlled ``pinching'' mechanism. In more detail, the waveguide is oriented parallel to the $x$-axis at a height $h$, spanning a total length equal to $D_x$, and is equipped with $M$ PAs positioned at predefined locations where only one PA is activated during each transmission interval. In addition, the inter-spacing between consecutive PAs is denoted as $\delta = D_x/M$, and the coordinates of the $k$-th PA are given by $\boldsymbol{\psi_p^{(k)}} = (x_k, 0, h)$, with $x_k = \frac{2k-1}{2}\,\delta$ and $k = 1, 2, \ldots, M$. Therefore, the wireless channel between the $k$-th PA and the user is modeled as 
{\small
\begin{equation}
h_1^{(k)} = \frac{\sqrt{\eta} e^{-j \frac{2\pi}{\lambda} |\boldsymbol{\psi_m} -\boldsymbol{\psi_p^{(k)}}|}}{|\boldsymbol{\psi_m} -\boldsymbol{\psi_p^{(k)}}|},
\end{equation}
}where $\eta = \frac{\lambda^2}{16 \pi^2}$ denotes the path loss at a reference distance of 1 m, $\lambda$ is the free-space wavelength, $j$ is the imaginary unit, and $|\cdot|$ denotes the Euclidean norm. Moreover, as the signal propagates along the dielectric waveguide, it undergoes a phase shift determined by the effective refractive index $n_{\mathrm{eff}}$, which defines the guided wavelength as $\lambda_g = \frac{\lambda}{n_{\mathrm{eff}}}$. Accordingly, the channel between the waveguide feed point at $\boldsymbol{\psi_0} = (0, 0, h)$ and the $k$-th PA position is expressed as $h_2^{(k)} =\sqrt{ e^{-\alpha |\boldsymbol{\psi_p^{(k)}} -\boldsymbol{\psi_0}|}} e^{-j \frac{2\pi}{\lambda_g} |\boldsymbol{\psi_p^{(k)}} -\boldsymbol{\psi_0}|}$, where $\boldsymbol{\psi_0} = (0, 0, h)$ denotes the location of the waveguide feeding point. Finally, the waveguide is characterized by an attenuation coefficient $\alpha\in [0, +\infty)$, which accounts for the intrinsic exponential power attenuation of the signal as it traverses the waveguide. Consequently, the received signal at the user when the $k$-th PA is used can be expressed as 
\begin{equation}
y_r = \sqrt{P_t} h_1^{(k)} h_2^{(k)} s + w_n,
\end{equation}
where $P_t$ is the transmit power, $s$ is the transmitted symbol with $\mathbb{E}[|s|^2] = 1$ and $\mathbb{E}[\cdot]$ denoting expectation, and $w_n$ is additive white Gaussian noise with zero mean and variance $\sigma^2$. Therefore, the received SNR corresponding to the $k$-th PA is written as
{\small
\begin{equation}\label{SNR1_discrete}
\gamma^{(k)} = \frac{\eta P_t e^{-\alpha |\boldsymbol{\psi_p^{(k)}} -\boldsymbol{\psi_0}|} \left| e^{-j \left( \frac{2\pi}{\lambda} |\boldsymbol{\psi_m} -\boldsymbol{\psi_p^{(k)}}| + \frac{2\pi}{\lambda_g} |\boldsymbol{\psi_p^{(k)}} -\boldsymbol{\psi_0}| \right)} \right|^2}{\sigma^2 |\boldsymbol{\psi_m} -\boldsymbol{\psi_p^{(k)}}|^2}.
\end{equation}
}
Finally, considering that $|e^{-jx}| = 1$, \eqref{SNR1_discrete} simplifies to
\begin{equation}\label{SNR2_discrete}
\gamma^{(k)} = \frac{\eta P_t e^{-\alpha |\boldsymbol{\psi_p^{(k)}} -\boldsymbol{\psi_0}|}}{\sigma^2 |\boldsymbol{\psi_m} - \boldsymbol{\psi_p^{(k)}}|^2}
= \frac{\eta P_t e^{-\alpha x_k}}{\sigma^2 \big( (x_m - x_k)^2 + y_m^2 + h^2 \big)}.
\end{equation}

\begin{figure}
    \centering
    \includegraphics[width=\columnwidth]{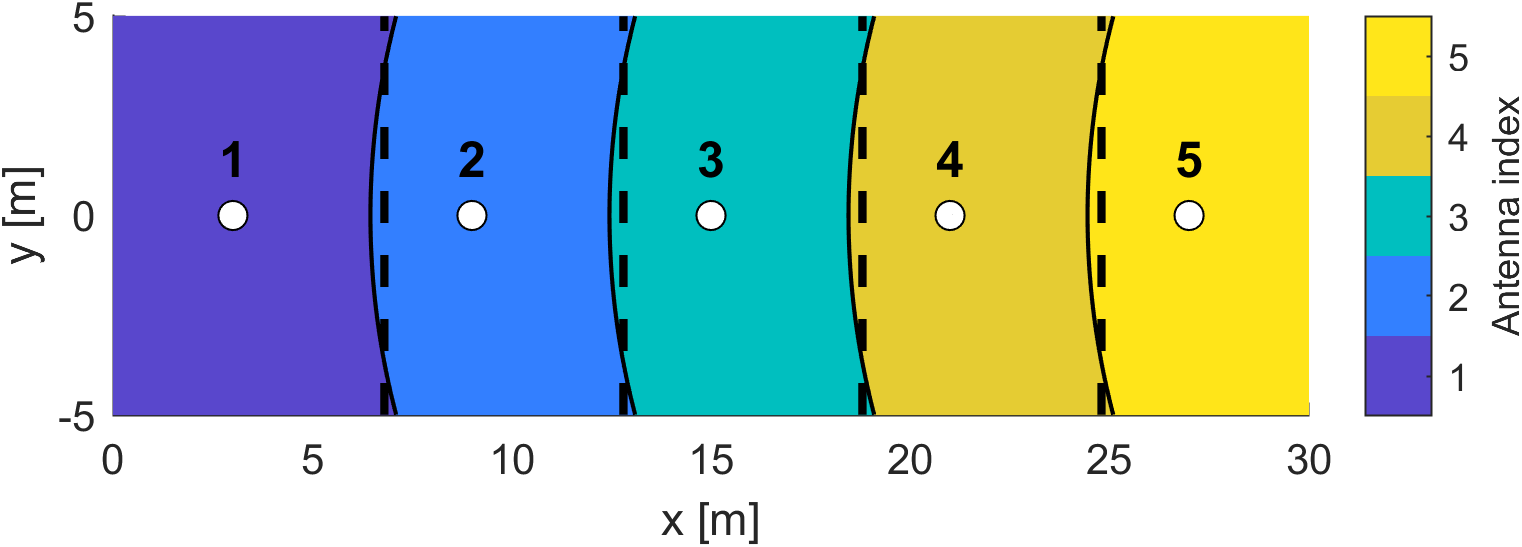}
    \caption{PA regions of two-state PAS with $M=5$, $D_x=30 \,m$ and $\alpha=0.05$.}
    \label{fig:snr_regions}
\end{figure}

To analyze the performance of the proposed two-state PAS, an accurate description of how each user location is assigned to a specific PA is required. In more detail, since the AP activates the PA that yields the highest received SNR, the deployment area is divided into regions within which each PA is optimal. Therefore, by equating the SNR expressions of adjacent PAs in \eqref{SNR2_discrete}, and due to the combined effect of free-space propagation and waveguide attenuation, the boundaries of these regions take the form of slightly curved arcs, as illustrated in Fig.~\ref{fig:snr_regions}, which can be expressed mathematically as
\begin{equation}
    \left(x - x_k - \frac{e^{\alpha\delta}\delta}{e^{\alpha\delta} - 1}\right)^{2} + y^{2} = \frac{e^{\alpha\delta}\delta^{2}}{\left(e^{\alpha\delta} - 1\right)^{2}} - h^{2}.
\end{equation}
This corresponds to a circle of radius $\rho =\sqrt{\frac{e^{\alpha\delta}\delta^{2}}{\left(e^{\alpha\delta} - 1\right)^{2}} - h^{2}}$,
and curvature $\kappa = 1/\rho$. However, for practical waveguide parameters, where $\alpha\in[0.01,0.1]$ and $\delta$ is on the order of a few meters, we have $\alpha\delta\ll1$, which implies $e^{\alpha \delta} \approx 1 + \alpha \delta$, and $\kappa \approx \frac{1}{\sqrt{\tfrac{1}{\alpha^{2}} - h^{2}}}$, yielding small curvature for typical indoor heights $h$. Consequently, the curvature can be neglected and the region served by the $k$-th PA is well approximated by a rectangle with horizontal limits extending $L_k$ to the left of $x_k$ and $R_k$ to the right of $x_k$, forming a rectangular region that is asymmetrical around the PA.


\begin{remark}
For the ideal case where $\alpha = 0$, the boundaries between adjacent PAs reduce to vertical lines and each PA region becomes an exact rectangle, which is symmetric around the corresponding PA. In this case, the optimal PA is the one closest to the user in the $x$-dimension. 
\end{remark}

\section{Performance Analysis}

In this section, we derive analytical expressions for key performance metrics, including the outage probability and the ergodic rate of a two-state PAS affected by waveguide attenuation. Specifically, the following analysis captures how the discrete PA positions and the waveguide attenuation shape the SNR, enabling a precise evaluation of the performance attained by practical two-state implementations. Finally, we introduce the PDE, which quantifies the rate deviation from the ideal continuous PAS and provides a direct measure of the impact of spatial discretization.

To design a two-state PAS, it is essential to evaluate the outage probability, since this metric can indicate the minimum number of PAs required to ensure reliable communication. In this direction, the following proposition provides a closed-form expression for the outage probability of the considered two-state PAS.
\begin{proposition}\label{Prop02}
The outage probability of the considered two-state PAS, where the $k$-th PA is located at $(x_k,0,h)$ and the user position is given by $(x_m,y_m,0)$, is given as
\begin{equation}
    P_o=\sum_{k=1}^{M} \left(\frac{L_k P_l(L_k)}{D_x}\,+\,\frac{R_k P_l(R_k)}{D_x} \right),
\end{equation}
where $P_l(\cdot)$ is given in Table \ref{TableOut}, with $A_{0,k}=\frac{C_{0,k}}{\gamma_{\mathrm{thr}}}-h^2$, $C_{0,k}=Ce^{-ax_k}$, $C=\frac{\eta P_t}{\sigma^2}$ and $\gamma_{\mathrm{thr}}$ is the threshold SNR value.
\begin{table*}[t]
\centering
\caption{Expressions and conditions for the probability $P_{l}(\Delta)$\label{TableOut}}
\renewcommand{\arraystretch}{1.6} 
\begin{tabular}{p{0.55\linewidth}||p{0.35\linewidth}}
\hline\hline
\textbf{Expression of $P_{l}(\Delta)$} & \textbf{Condition} \\
\hline\hline

$\large 1$ 
& 
$\large A_{0,k} \leq 0$
\\\hline

$\large 1 - \frac{\pi A_{0,k}}{2 D_y \Delta}$ 
& 
$\large 0 \leq A_{0,k} \leq \min\!\left(\frac{D_y^{2}}{4},\, \Delta^{2}\right)$
\\\hline

$\large 1 - \frac{A_{0,k}}{D_y \Delta}\, 
\arcsin\!\left(\frac{\Delta}{\sqrt{A_{0,k}}}\right)
- \frac{\sqrt{A_{0,k}-\Delta^{2}}}{D_y}$ 
& 
$\large \Delta^{2} \leq A_{0,k} \leq \frac{D_y^{2}}{4}$
\\\hline

$\large 1 - \frac{A_{0,k}}{D_y \Delta}\,
\arcsin\!\left(\frac{D_y}{2\sqrt{A_{0,k}}}\right)
- \frac{\sqrt{A_{0,k}-\frac{D_y^{2}}{4}}}{2\Delta}$ 
& 
$\large \frac{D_y^{2}}{4} \leq A_{0,k} \leq \Delta^{2}$
\\\hline

$\large 
1 - \frac{A_{0,k}}{D_y \Delta}\!
\left(
\arcsin\!\left(\frac{\Delta}{\sqrt{A_{0,k}}}\right)
-
\arcsin\!\left(\frac{\sqrt{4A_{0,k}-D_y^{2}}}{2\sqrt{A_{0,k}}}\right)
\right)
- \frac{1}{4\Delta}\sqrt{4A_{0,k}-D_y^{2}}
- \frac{1}{D_y}\sqrt{A_{0,k}-\Delta^{2}}
$
&
$\large\max\!\left(\Delta^{2},\, \frac{D_y^{2}}{4}\right)
\leq A_{0,k} \leq
\Delta^{2} + \frac{D_y^{2}}{4}
$
\\\hline

$\large 0$ 
& 
$\large A_{0,k} \geq \Delta^{2} + \frac{D_y^{2}}{4}$
\\\hline\hline
\end{tabular}
\end{table*}
\end{proposition}
\begin{IEEEproof}
By taking into account that $x_k$ is a discrete random variable, the outage probability can be written as
\begin{equation}
\begin{split}  
P_o
=
&\sum_{k=1}^{M} 
\Pr\{ \text{the $k$-th PA is selected}\}  \\
& \times \Pr\{\gamma^{(k)} \leq \gamma_{\mathrm{thr}} \mid \text{the $k$-th PA is selected}\}.
\end{split}
\end{equation}
Under the rectangular approximation, the region associated with the $k$-th PA is generally asymmetric around $x_k$, since $L_k \neq R_k$. Therefore, by splitting this region into two sub-rectangles, namely a left part with $\varepsilon \in [-L_k,0]$ and a right part with $\varepsilon \in [0,R_k]$, the conditional outage event can be expressed as the sum of the outage probabilities in each sub-region. Because $x_m$ is uniformly distributed over $[0,D_x]$, the probabilities that the user lies in the left and right sub-rectangles are $\frac{L_k}{D_x}$ and $\frac{R_k}{D_x}$, respectively. Therefore, the outage probability of the considered two-state PAS equals to
\begin{equation}
P_o
=
\sum_{k=1}^{M}
\left(
\frac{L_k}{D_x} P_l(L_k)
+
\frac{R_k}{D_x} P_l(R_k)
\right),
\end{equation}
where $P_l(\Delta)$ denotes the conditional outage probability when the user lies in a sub-rectangle of width $\Delta$.

To determine $P_l(\Delta)$, we focus on a sub-rectangle of width $\Delta$ associated with a given PA, where the horizontal displacement $\varepsilon = x_m - x_k$ is uniformly distributed over $[0,\Delta]$. Hence, using \eqref{SNR2_discrete}, the conditional outage probability in a sub-rectangle of width $\Delta$ is written as
\begin{equation}
P_l(\Delta)
=
\Pr\!\left\{ y^2_m \geq A_{0,k}-\varepsilon^2
\right\}.
\end{equation}
Therefore, to derive $P_l(\Delta)$, we need to identify the feasible region for $y_m$ based on the condition $y^2_m \geq A_{0,k}-\varepsilon^2$, and then integrate over the joint probability density function of $\varepsilon$ and $y_m$, which is equal to $f_{\varepsilon,y_m}(\varepsilon,y_m) = \frac{1}{\Delta D_y}$.
\begin{itemize}
    \item If $A_{0,k}\leq 0$, then $P_l(\Delta)=1$,
    \item If $ A_{0,k} \geq \Delta^{2} + \frac{D_y^{2}}{4}$, then $P_l(\Delta)=0$,
    \item If $ 0 \leq A_{0,k} \leq \Delta^{2} + \frac{D_y^{2}}{4}$ then $y_m \in [-\frac{D_y}{2}, - \sqrt{A_{0,k}-\varepsilon^2}] \bigcup [\sqrt{A_{0,k}-\varepsilon^2}, \frac{D_y}{2}]$.
\end{itemize}
For the third case, where $0 \leq A_{0,k} \leq \Delta^{2} + \frac{D_y^{2}}{4}$, the conditional outage probability $P_l(\Delta)$ can be expressed mathematically as
\begin{equation}
\small
\label{eq:Pl_general_integral}
\begin{aligned}
P_l(\Delta)
&=
\frac{1}{\Delta D_y}
\int_{A_1}^{A_2}
\left[
\int_{-B_2(\varepsilon)}^{-B_1(\varepsilon)} dy_m
+
\int_{B_1(\varepsilon)}^{B_2(\varepsilon)} dy_m
\right]
d\varepsilon
\\[1mm]
&\quad+
\frac{1}{\Delta D_y}
\int_{A_2}^{A_3}
\int_{-\frac{D_y}{2}}^{\frac{D_y}{2}}
dy_m \, d\varepsilon ,
\end{aligned}
\end{equation}
where the limits $A_1$, $A_2$, and $A_3$, as well as $B_1(\varepsilon)$ and $B_2(\varepsilon)$, are determined by the relative values of $A_{0,k}$, $\Delta^2$, and $\tfrac{D_y^2}{4}$. Thus, four distinct parameter regimes arise,
\begin{itemize}
    \item[(i)] $A_{0,k} \leq \min\!\left(\Delta^{2},\frac{D_y^{2}}{4}\right)$,
    \item[(ii)] $\Delta^{2} \leq A_{0,k} \leq \frac{D_y^{2}}{4}$,
    \item[(iii)] \(\frac{D_y^{2}}{4} \leq A_{0,k} \leq \Delta^{2}\),
    \item[(iv)] $\max\!\left(\Delta^{2},\frac{D_y^{2}}{4}\right) \leq A_{0,k} \leq \Delta^{2} + \frac{D_y^{2}}{4}$,
\end{itemize}
based on which the integration limits of \eqref{eq:Pl_general_integral} are adjusted accordingly. Therefore, by evaluating the resulting integrals for each regime as shown in Appendix \ref{App:1}, we obtain the piecewise closed-form expressions of $P_l(\Delta)$ summarized in Table~\ref{TableOut}, which concludes the proof.
\end{IEEEproof}

To further characterize the performance of the considered two-state PAS, it is essential to derive its ergodic data rate, which captures the average spectral efficiency over the spatial distribution of the user. In more detail, the ergodic rate depends on the full SNR profile within each PA-associated region and thus inherits the spatial discreteness of the available PA positions, revealing a direct coupling between the geometry of the PAS deployment and the achievable communication throughput. In this direction, the following proposition provides a closed-form expression for the ergodic data rate, highlighting the impact of the number of PAs and the room dimensions on the overall performance.

\begin{proposition}
Considering that $\varepsilon \sim \mathcal{U}\!\left[0,\Delta\right]$ and $y_m \sim \mathcal{U}\!\left[-\frac{D_y}{2},\frac{D_y}{2}\right]$, the ergodic rate of the considered two-state PAS can be expressed as
\begin{equation}\label{R_avg_f}
\overline{R} = \sum_{k=1}^{M} \left(\frac{L_k C_l(L_k)}{D_x}\,+\,\frac{R_k C_l(R_k)}{D_x} \right),
\end{equation}
where 
\begin{equation}\label{R_avg}
\begin{split}
C_l(\Delta)= &\frac{2}{\Delta D_y \ln 2}\,\Big(I_i(C_{0,k}+h^2) + I_j(C_{0,k}+h^2) \\&- I_i(h^2) - I_j(h^2)\Big),
\end{split}
\end{equation} 
and $I_i(\cdot)$ and $I_j(\cdot)$ are given in \eqref{Iix} and \eqref{Ijx} at the top of this page, respectively, and $\mathrm{Ti}_2(z)=\tfrac{\mathrm{Li}_2(j z)-\mathrm{Li}_2(-j z)}{2j}$ denotes the arctangent integral function, with $\mathrm{Li}_2(\cdot)$ representing the dilogarithm function.
\end{proposition}

\begin{figure*}[t]
\begin{equation}\label{Iix}
\small
I_i(x)
= \frac{\Delta D_y}{2}\,
    \ln\!\left(\frac{D_y^2}{4}+\Delta^2+x\right)
   -\Delta D_y
   +2\Delta\,\sqrt{\Delta^2+x}\;
     \tan^{-1}\!\left(
       \frac{D_y}{2\sqrt{x+\Delta^2}}
     \right)
\end{equation}
\hrule
\begin{equation}\label{Ijx}
\small
\begin{split}
I_j(x)
&=\Bigg[
     \frac{D_y}{2}\sqrt{x+\frac{D_y^2}{4}}
     + x
       \ln\!\left(\frac{D_y}{2}+\sqrt{x+\frac{D_y^2}{4}}\right)
  \Bigg]
  \tan^{-1}\!\left(
     \frac{\Delta}{\sqrt{x+\frac{D_y^2}{4}}}
  \right) -x\ln\!\big(\sqrt{x}\big)
     \tan^{-1}\!\left(\frac{\Delta}{\sqrt{x}}\right)+\frac{\Delta D_y}{2}
  \\
&\quad
  -\Delta\sqrt{x+\Delta^2}\;
     \tan^{-1}\!\left(
        \frac{D_y}{2\sqrt{x+\Delta^2}}
     \right) +x\Bigg[
     \Big(
        \ln\!\big(\sqrt{x}\big)
        + \operatorname{asinh}\!\Big(\frac{D_y}{2\sqrt{x}}\Big)
     \Big)
     \tan^{-1}\!\left(\frac{\sqrt{x+\frac{D_y^2}{4}}}{\Delta}\right)\\
&\qquad
      - \ln\!\big(\sqrt{x}\big)
       \tan^{-1}\!\Big(\frac{\sqrt{x}}{\Delta}\Big)
     - \frac{\pi}{2}\,
       \operatorname{asinh}\!\Big(\frac{D_y}{2\sqrt{x}}\Big)
     -\,\mathrm{Ti}_2\Big(
         e^{\operatorname{asinh}\!\left(\frac{D_y}{2\sqrt{x}}\right)}
         \frac{\Delta}{\sqrt{x}}\,
         \big(\sqrt{1+\tfrac{x}{\Delta^2}}-1\big)
       \Big) 
    \\
&\qquad
       +\,\mathrm{Ti}_2\Big(
         \frac{\Delta}{\sqrt{x}}\,
         \big(\sqrt{1+\tfrac{x}{\Delta^2}}-1\big)
       \Big) -\,\mathrm{Ti}_2\Big(
         -e^{\operatorname{asinh}\!\left(\frac{D_y}{2\sqrt{x}}\right)}
         \frac{\Delta}{\sqrt{x}}\,
         \big(\sqrt{1+\tfrac{x}{\Delta^2}}+1\big)
       \Big) +\,\mathrm{Ti}_2\Big(
         -\frac{\Delta}{\sqrt{x}}\,
         \big(\sqrt{1+\tfrac{x}{\Delta^2}}+1\big)
       \Big)
        \Bigg]
\end{split}
\end{equation}
\hrule
\end{figure*}

\begin{IEEEproof} 
Since the active PA is selected according to the user’s horizontal position, the ergodic rate can be written as the weighted sum of the conditional ergodic rates over the left and right sub-rectangles associated with each PA. Hence, the average rate of the considered two-state PAS can be expressed as in \eqref{R_avg_f}, where $C_l(\Delta)$ denotes the conditional ergodic rate when the user lies uniformly within a sub-rectangle of width $\Delta$. Therefore, by taking into account \eqref{SNR2_discrete}, $C_l(\Delta)$ can be expressed as
\begin{equation}\label{c1}
C_l(\Delta)=\mathbb{E}\!\left[\log_2\!\left(1+\frac{C_{0,k}}{h^2+\varepsilon^2+y_m^2}\right)\right] .
\end{equation}
Since \eqref{c1} is an even function with respect to $y_m$, \eqref{c1} can be written in integral form as follows
\begin{equation}
\small
    C_l(\Delta)=\frac{2}{\Delta D_y\ln2} \int_0^{\frac{D_y}{2}} \int_0^{\Delta} \ln \left(1+ \frac{C_{0,k}}{\varepsilon^2 + y_m^2 + h^2 } \right) d\varepsilon \, dy_m,
\end{equation}
which, after some algebraic manipulations, can be rewritten as in \eqref{R_avg}, where $I_i(\cdot)$ and $I_j(\cdot)$ are equal to
\begin{equation}
    I_i(x) = \int_0^{\frac{D_y}{2}} \Delta \ln\left( \Delta^2 + x +y^2_m\right) dy_m,
\end{equation}
and
\begin{equation}\label{Ij_13}
    I_j(x) = \int_0^{\frac{D_y}{2}} 2 \sqrt{x+ y^2_m} \tan^{-1} \! \left( \frac{\Delta}{\sqrt{x+ y^2_m}}\right)dy_m.
\end{equation}
Thus, by following similar steps as shown in Appendix \ref{App:A}, we obtain \eqref{Iix} and \eqref{Ijx}, and by substituting them in \eqref{R_avg}, \eqref{R_avg_f} is derived, which concludes the proof.
\end{IEEEproof}

\begin{remark}
For the ideal case where $\alpha = 0$, it holds that $C_{0,k} = C$ for all $k$. As a result, $A_{0,k} = \frac{C}{\gamma_{\mathrm{thr}}} - h^2$ is independent of $k$. Moreover, in the absence of waveguide attenuation, the PA regions become symmetric around $x_k$, which implies $L_k = R_k = \tfrac{\delta}{2}$ and therefore $P_l(L_k) = P_l(R_k) = P_l\!\left(\tfrac{\delta}{2}\right)$, and $C_l(L_k) = C_l(R_k) = C_l\!\left(\tfrac{\delta}{2}\right)$.
\end{remark}

Based on the derived closed-form expression of the ergodic rate, we further define the PDE as
\begin{equation}\label{eq:ratio}
\overline{\eta}_r=\frac{\overline{R}}{R_{\mathrm{c}}},
\end{equation}
where $R_{\mathrm{c}}$ denotes the ergodic data rate of an ideal continuous PAS, in which a PA can be placed at any arbitrary point along the waveguide, and it is deployed at the optimal position $x_p^{\star}$ whose closed-form expression is provided in \cite{TyrovolasPASS2025}. Thus, the PDE quantifies the rate loss that arises when only a finite number of pinching positions are available, thereby completing the analytical characterization of the proposed two-state PAS.

\begin{remark}\label{Rem:PDE_not_one}
Due to waveguide attenuation, the optimal PA position in a continuous PAS does not coincide with the user’s horizontal position but lies at a point strictly before it, as characterized by $x_p^\star$ in \cite{TyrovolasPASS2025}. In this direction, as the density of available PAs increases, the discrete candidate locations initially approach $x_p^\star$, which leads to improved performance. However, further densification causes the available PA positions to move closer to the user location and away from $x_p^\star$.
This behavior indicates that, in the presence of waveguide attenuation, a two-state PAS may realize a more effective serving strategy than always placing the PA directly above the user. 
\end{remark}



\section{Numerical Results}
In this section, we evaluate the performance of the examined two-state PAS and the accuracy and validity of the derived analytical expressions. For consistency, the system parameters are chosen as in \cite{Ding2024TCOM}, where the noise power $\sigma^2$ is $-90$~dBm, the carrier frequency is $f_c=28$~GHz, and the effective refractive index is $n_{\mathrm{eff}}=1.4$. Moreover, unless stated otherwise, the attenuation factor is set to $\alpha=0.05$, the SNR threshold is $\gamma_{\mathrm{thr}}=20$ dB, the waveguide height is $h=3$~m, and the deployment area has dimension $D_y=10$~m. Furthermore, for each PA, the horizontal serving limits $L_k$ and $R_k$ are obtained through a one-dimensional numerical optimization, yielding optimized rectangular limits that closely match the actual serving regions. Finally, to validate the theoretical results, Monte Carlo simulations with $10^6$ random realizations are performed.

\begin{figure}[h!]
    \centering
    \begin{subfigure}{\linewidth}
        \centering
        \begin{tikzpicture}
        \begin{semilogyaxis}[
            width=0.99\linewidth,
            xlabel = {$\gamma_t$ (dB)},
            ylabel = {$P_o$},
            xmin = 90, xmax = 110,
            ymin = 1e-5, ymax = 1,
            xtick = {90,95,...,110},
            grid = major,
            legend image post style={xscale=0.9, every mark/.append style={solid}},
            legend cell align = {left},
            legend style={
                at={(1,1)},
                anchor=north east,
                font = \tiny
            }
        ]

        \addplot[
            red,
            no marks,
            line width = 1pt,
            solid
        ]
        table {figures_tr/Fig1_10_1.dat};
        \addlegendentry{$M=1$}
        
        \addplot[
            black,
            no marks,
            line width = 1pt,
            solid
        ]
        table {figures_tr/Fig1_10_2.dat};
        \addlegendentry{$M=2$}
        
        \addplot[
            blue,
            no marks,
            line width = 1pt,
            solid
        ]
        table {figures_tr/Fig1_10_10.dat};
        \addlegendentry{$M=10$}

        \addplot[
            black,
            only marks,
            mark=triangle*,
            mark size=3,
            mark indices={2},
        ]
        table {figures_tr/Fig1_10_2.dat};
        \addlegendentry{Simulation Results}
                
        \addplot[
            black,
            only marks,
            mark=triangle*,
            mark size=3,
            mark indices={0, 2, 5, 9, 13, 18, 25, 33, 43, 56, 65, 69, 71},
        ]
        table {figures_tr/Fig1_10_2.dat};
        
        \addplot[
            red,
            only marks,
            mark=triangle*,
            mark size = 3,
            mark indices={0, 26, 60, 102, 155, 222, 307, 415, 551, 732, 870, 920, 938},
        ]
        table {figures_tr/Fig1_10_1.dat};

        \addplot[
            blue,
            only marks,
            mark=triangle*,
            mark size = 3,
            mark indices={0, 11, 25, 42, 62, 87, 118, 156, 202, 259, 290, 320, 327},
        ]
        table {figures_tr/Fig1_10_10.dat};

        \end{semilogyaxis}
        \end{tikzpicture}
        \caption{}
    \end{subfigure}

    \begin{subfigure}{\linewidth}
        \centering
        \begin{tikzpicture}
        \begin{semilogyaxis}[
            width=0.99\linewidth,
            xlabel = {$\gamma_t$ (dB)},
            ylabel = {$P_o$},
            xmin = 90, xmax = 110,
            ymin = 1e-5, ymax = 1,
            xtick = {90,95,...,110},
            grid = major,
            legend image post style={xscale=0.9, every mark/.append style={solid}},
            legend cell align = {left},
            legend style={
                at={(0,0)},
                anchor=south west,
                font = \tiny
            }
        ]

        \addplot[
            red,
            no marks,
            line width = 1pt,
            solid
        ]
        table {figures_tr/Fig1_30_1_reduced.dat};
        \addlegendentry{$M=1$}
        
        \addplot[
            black,
            no marks,
            line width = 1pt,
            solid
        ]
        table {figures_tr/Fig1_30_2_reduced.dat};
        \addlegendentry{$M=2$}
        
        \addplot[
            blue,
            no marks,
            line width = 1pt,
            solid
        ]
        table {figures_tr/Fig1_30_10_reduced.dat};
        \addlegendentry{$M=10$}

        \addplot[
            black,
            only marks,
            mark=triangle*,
            mark size=3,
            mark indices={2},
        ]
        table {figures_tr/Fig1_30_2_reduced.dat};
        \addlegendentry{Simulation Results}
                
        \addplot[
            black,
            only marks,
            mark=triangle*,
            mark size=3,
            mark indices={0, 7, 17, 33, 58, 96, 154, 244, 352,480, 595, 700, 787, 856, 890, 912, 924},
        ]
        table {figures_tr/Fig1_30_2_reduced.dat};
        
        \addplot[
            red,
            only marks,
            mark=triangle*,
            mark size = 3,
            mark indices={0, 14, 34, 66, 116, 192, 309, 490, 768, 1196, 1580, 1720, 1790, 1830, 1848, 1856},
        ]
        table {figures_tr/Fig1_30_1_reduced.dat};

        \addplot[
            blue,
            only marks,
            mark=triangle*,
            mark size = 3,
            mark indices={0, 7, 18, 35, 61, 101, 163, 258, 405, 631, 833, 907, 944, 965, 978},
        ]
        table {figures_tr/Fig1_30_10_reduced.dat};

        \end{semilogyaxis}
        \end{tikzpicture}
        \caption{}
    \end{subfigure}

    \caption{Outage probability versus $\gamma_t$ for a two-state PAS for various $M$ values and a) $D_x = 10$ m, and b) $D_x = 30$ m.}
    \label{Fig3}
\end{figure}
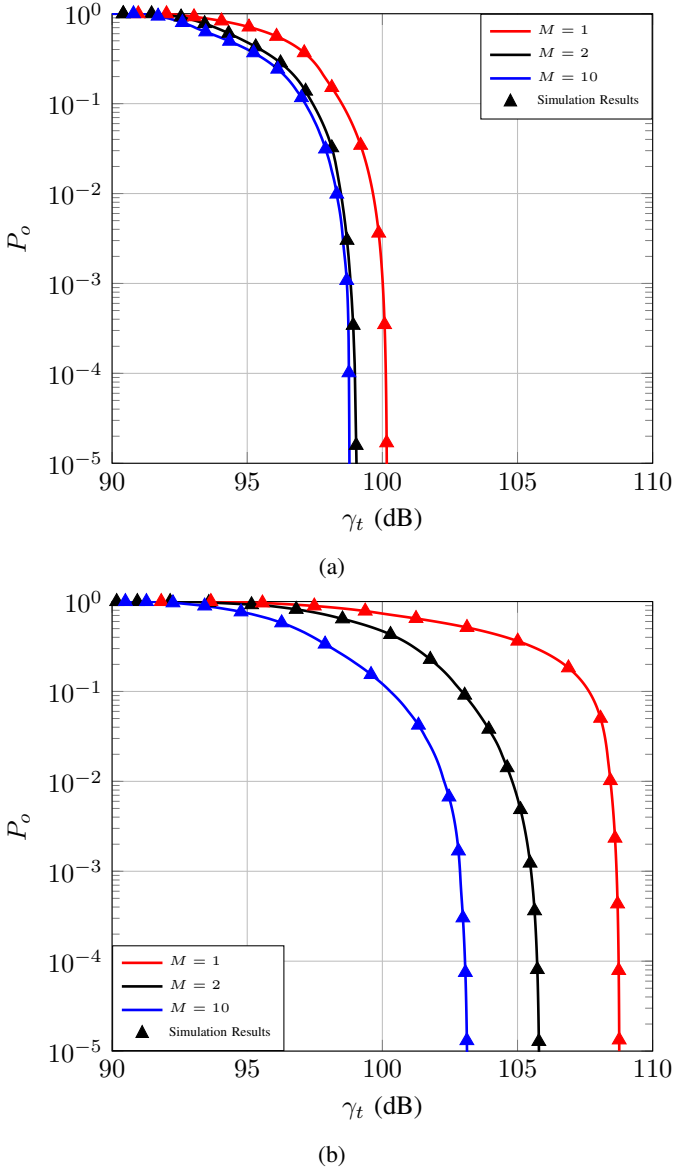

Figs.~\ref{Fig3}a and \ref{Fig3}b illustrate the outage probability of the two-state PAS as a function of the SNR threshold $\gamma_t$ for different numbers of PAs, where Fig.~\ref{Fig3}a corresponds to $D_x=10$~m and Fig.~\ref{Fig3}b to $D_x=30$~m. In both cases, the analytical curves closely match the Monte Carlo simulation results, validating the accuracy of the derived outage expressions. As expected, increasing the number of PAs improves the outage performance by increasing the likelihood that the user is served by a more favorable antenna position. However, Fig.~\ref{Fig3}a shows that for smaller deployment widths the outage performance quickly saturates, since the curves for $M=2$ and $M=10$ are almost identical, indicating that additional PAs provide negligible gains. In contrast, for larger deployment widths, as shown in Fig.~\ref{Fig3}b, the performance improvement persists over a wider range of $M$, although similar saturation behavior is expected to occur for sufficiently large numbers of PAs. These results highlight that the two-state PAS can achieve near-optimal outage performance with a limited number of PAs, while larger environments primarily shift the saturation point to higher PA densities rather than eliminating it.

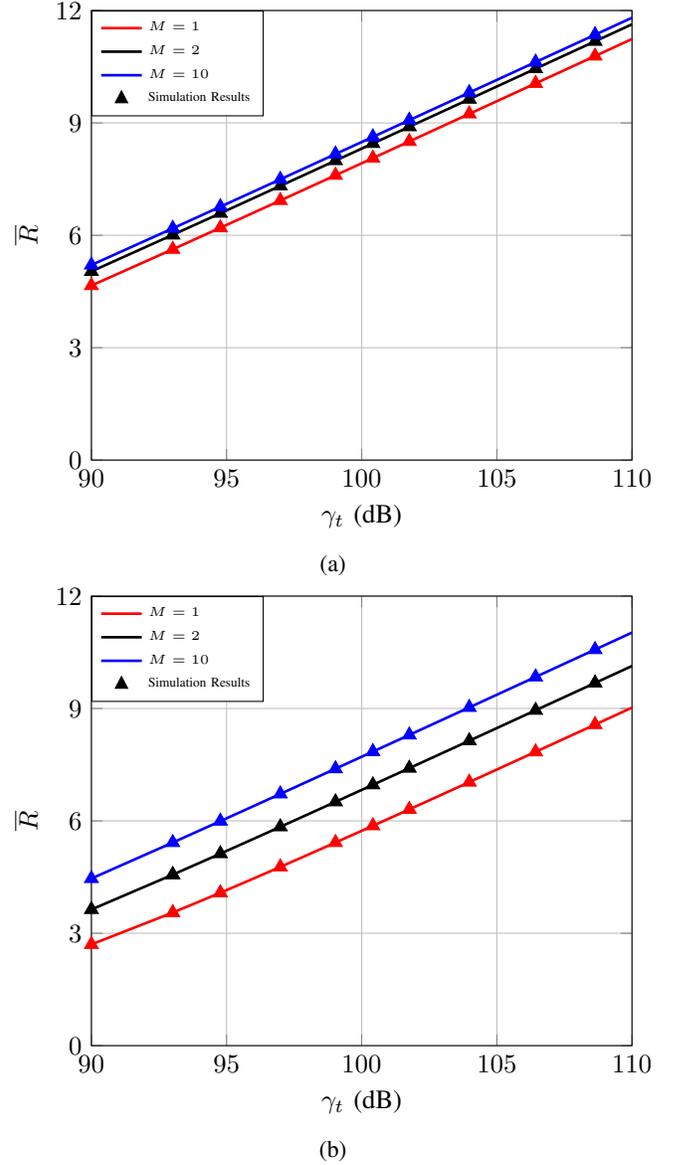
\begin{figure}[h!]
    \centering
    \begin{subfigure}{\linewidth}
        \centering
        \begin{tikzpicture}
        \begin{axis}[
            width=0.99\linewidth,
            xlabel = {$\gamma_t$ (dB)},
            ylabel = {$\overline{R}$},
            xmin = 90, xmax = 110,
            ymin = 0,  ymax = 12,
            xtick = {90,95,...,110},
            ytick = {0,3,6,9,12},
            grid = major,
            legend image post style={xscale=0.9, every mark/.append style={solid}},
            legend cell align = {left},
            legend style={
                at={(0,1)},
                anchor=north west,
                font = \tiny
            }
        ]

        \addplot[
            red,
            no marks,
            line width = 1pt,
            solid
        ]
        table {figures_tr/Fig2_10_1.dat};
        \addlegendentry{$M=1$}

        \addplot[
            black,
            no marks,
            line width = 1pt,
            solid
        ]
        table {figures_tr/Fig2_10_2.dat};
        \addlegendentry{$M=2$}

        \addplot[
            blue,
            no marks,
            line width = 1pt,
            solid
        ]
        table {figures_tr/Fig2_10_10.dat};
        \addlegendentry{$M=10$}

        \addplot[
            black,
            only marks,
            mark=triangle*,
            mark size=3,
            mark indices={2},
        ]
        table {figures_tr/Fig2_10_2.dat};
        \addlegendentry{Simulation Results}

        \addplot[
            black,
            only marks,
            mark=triangle*,
            mark size=3,
            mark indices={0, 1, 2, 3, 5, 8, 11, 15, 25, 44, 73},
        ]
        table {figures_tr/Fig2_10_2.dat};

        \addplot[
            red,
            only marks,
            mark=triangle*,
            mark size = 3,
            mark indices={0, 1, 2, 3, 5, 8, 11, 15, 25, 44, 73},
        ]
        table {figures_tr/Fig2_10_1.dat};

        \addplot[
            blue,
            only marks,
            mark=triangle*,
            mark size = 3,
            mark indices={0, 1, 2, 3, 5, 8, 11, 15, 25, 44, 73},
        ]
        table {figures_tr/Fig2_10_10.dat};

        \end{axis}
        \end{tikzpicture}
        \caption{}
    \end{subfigure}

    \begin{subfigure}{\linewidth}
        \centering
        \begin{tikzpicture}
        \begin{axis}[
            width=0.99\linewidth,
            xlabel = {$\gamma_t$ (dB)},
            ylabel = {$\overline{R}$},
            xmin = 90, xmax = 110,
            ymin = 0,  ymax = 12,
            xtick = {90,95,...,110},
            ytick = {0,3,6,9,12},
            grid = major,
            legend image post style={xscale=0.9, every mark/.append style={solid}},
            legend cell align = {left},
            legend style={
                at={(0,1)},
                anchor=north west,
                font = \tiny
            }
        ]

        \addplot[
            red,
            no marks,
            line width = 1pt,
            solid
        ]
        table {figures_tr/Fig2_30_1.dat};
        \addlegendentry{$M=1$}

        \addplot[
            black,
            no marks,
            line width = 1pt,
            solid
        ]
        table {figures_tr/Fig2_30_2.dat};
        \addlegendentry{$M=2$}

        \addplot[
            blue,
            no marks,
            line width = 1pt,
            solid
        ]
        table {figures_tr/Fig2_30_10.dat};
        \addlegendentry{$M=10$}

        \addplot[
            black,
            only marks,
            mark=triangle*,
            mark size=3,
            mark indices={2},
        ]
        table {figures_tr/Fig2_30_2.dat};
        \addlegendentry{Simulation Results}

        \addplot[
            black,
            only marks,
            mark=triangle*,
            mark size=3,
            mark indices={0, 1, 2, 3, 5, 8, 11, 15, 25, 44, 73},
        ]
        table {figures_tr/Fig2_30_2.dat};

        \addplot[
            red,
            only marks,
            mark=triangle*,
            mark size = 3,
            mark indices={0, 1, 2, 3, 5, 8, 11, 15, 25, 44, 73},
        ]
        table {figures_tr/Fig2_30_1.dat};

        \addplot[
            blue,
            only marks,
            mark=triangle*,
            mark size = 3,
            mark indices={0, 1, 2, 3, 5, 8, 11, 15, 25, 44, 73},
        ]
        table {figures_tr/Fig2_30_10.dat};

        \end{axis}
        \end{tikzpicture}
        \caption{}
    \end{subfigure}

    \caption{Ergodic data rate versus $\gamma_t$ for a two-state PAS for various $M$ values and a) $D_x=10$ m, and b) $D_x=30$ m.}
    \label{Fig4}
\end{figure}

In Figs.~\ref{Fig4}a and \ref{Fig4}b, the ergodic data rate of the two-state PAS is illustrated as a function of $\gamma_t$ for different numbers of PAs, where Fig.~\ref{Fig4}a corresponds to a deployment width of $D_x=10$~m and Fig.~\ref{Fig4}b to $D_x=30$~m. In both cases, the theoretical curves obtained from the derived closed-form expressions closely follow the Monte Carlo simulation results, validating the accuracy of the developed analytical framework. As shown in Fig.~\ref{Fig4}a, for smaller deployment widths, increasing the number of PAs results in only marginal improvements in the achievable ergodic data rate, indicating an early saturation of spatial gains, which is consistent with the outage performance observed for compact environments. In contrast, as shown in Fig.~\ref{Fig4}b, when $D_x$ increases, the separation between the curves becomes more pronounced and the achievable data rate improves noticeably with the number of PAs, highlighting that additional PAs become increasingly beneficial in wider deployments where larger spatial separation mitigates propagation losses more effectively and enhances link reliability.

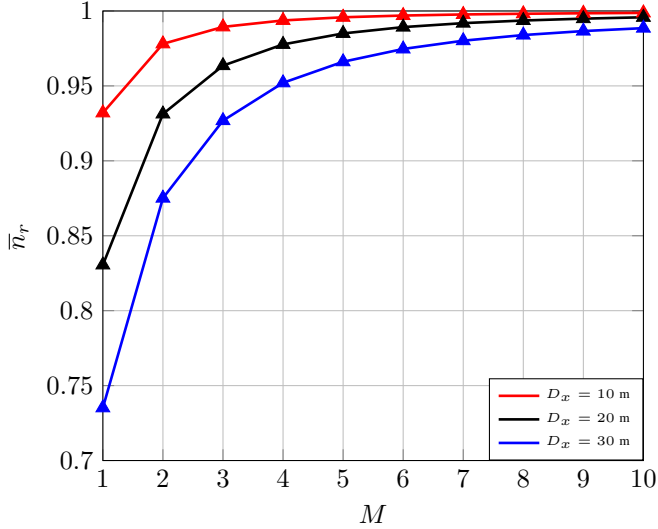
\begin{figure}
    \centering
    \begin{subfigure}{\linewidth}
        \centering
        \begin{tikzpicture}
        \begin{axis}[
            width=0.99\linewidth,
            xlabel = {$M$},
            ylabel = {$\overline{n}_r$},
            xmin = 1, xmax = 10,
            ymin = 0.7, ymax = 1,
            xtick = {1,2,3,4,5,6,7,8,9,10},
            ytick = {0.7,0.75,0.8,0.85,0.9,0.95,1},
            grid = major,
            legend image post style={xscale=0.9, every mark/.append style={solid}},
            legend cell align = {left},
            legend style={
                at={(1,0)},
                anchor=south east,
                font = \tiny
            }
        ]

        \addplot[
            red,
            no marks,
            line width = 1pt,
            solid
        ]
        table {figures_tr/Fig3_10_005.dat};
        \addlegendentry{$D_x=10$ m}

        \addplot[
            black,
            no marks,
            line width = 1pt,
            solid
        ]
        table {figures_tr/Fig3_20_005.dat};
        \addlegendentry{$D_x=20$ m}

        \addplot[
            blue,
            no marks,
            line width = 1pt,
            solid
        ]
        table {figures_tr/Fig3_30_005.dat};
        \addlegendentry{$D_x=30$ m}

        \addplot[
            red,
            only marks,
            mark=triangle*,
            mark size = 3,
            mark repeat=1,
        ]
        table {figures_tr/Fig3_10_005.dat};

        \addplot[
            black,
            only marks,
            mark=triangle*,
            mark size = 3,
            mark repeat=1,
        ]
        table {figures_tr/Fig3_20_005.dat};

        \addplot[
            blue,
            only marks,
            mark=triangle*,
            mark size = 3,
            mark repeat=1,
        ]
        table {figures_tr/Fig3_30_005.dat};

        \end{axis}
        \end{tikzpicture}
    \end{subfigure}
    \caption{PDE versus $M$ for $\gamma_t = 90$ dB.}
    \label{Fig5}
\end{figure}

Fig.~\ref{Fig5} illustrates the PDE as a function of the number of PAs \(M\) for \(\gamma_t = 90\)~dB. As expected, the PDE increases with \(M\), showing how the discrete PA configuration approaches the continuous PAS as the number of available PAs grows. In more detail, for compact deployments such as \(D_x = 10\)~m, the efficiency rapidly saturates, exceeding \(95\%\) with only two PAs, which reveals that in small-scale environments the discretization of the pinching process introduces only negligible performance degradation. However, as the deployment width increases, the PDE becomes more sensitive to the number of PAs, and a larger \(M\) is required to achieve the same performance level, with approximately three PAs needed for \(D_x = 20\)~m and four PAs for \(D_x = 30\)~m, which arises due to the increased PA spacing associated with wider deployments. Finally, the saturation of the PDE below unity indicates that even as the PA locations become denser, the two-state PAS cannot fully replicate the optimal PA placement of a continuous PAS due to waveguide attenuation, thus, indicating that perfect PDE cannot be attained in practice.

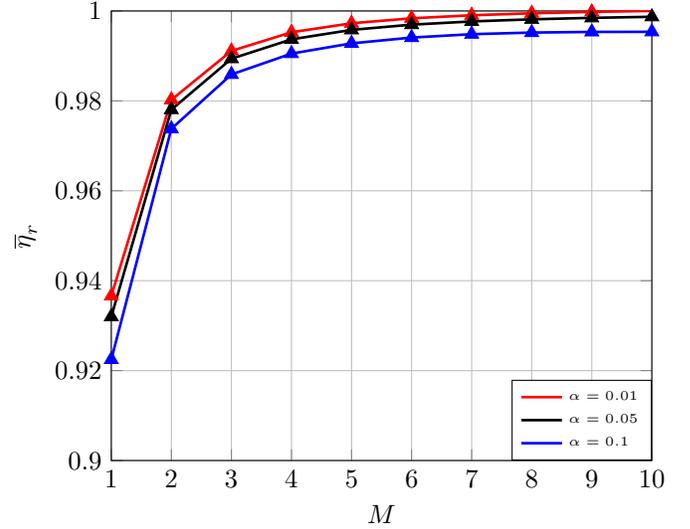
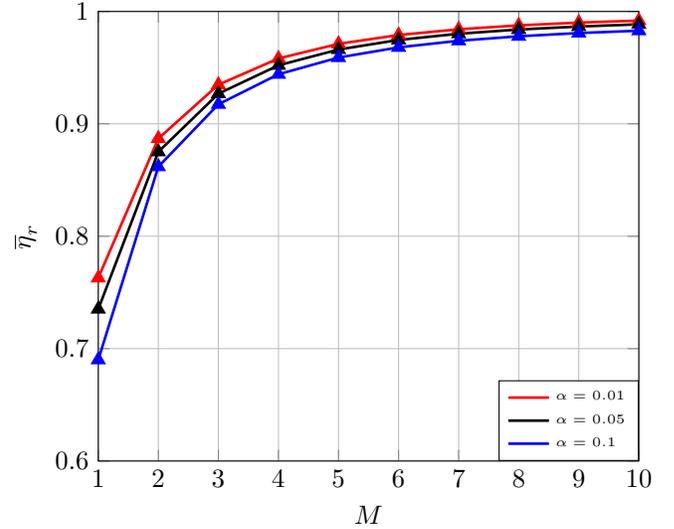
\begin{figure}[h!]
    \centering

    \begin{subfigure}{\linewidth}
        \centering
        \begin{tikzpicture}
        \begin{axis}[
            width=0.99\linewidth,
            xlabel = {$M$},
            ylabel = {$\overline{\eta}_r$},
            xmin = 1, xmax = 10,
            ymin = 0.9, ymax = 1,
            xtick = {1,2,3,4,5,6,7,8,9,10},
            ytick = {0.9,0.92,0.94,0.96,0.98,1},
            grid = major,
            legend image post style={xscale=0.9, every mark/.append style={solid}},
            legend cell align = {left},
            legend style={
                at={(1,0)},
                anchor=south east,
                font = \tiny
            }
        ]

        \addplot[
            red,
            no marks,
            line width = 1pt,
            solid
        ]
        table {figures_tr/Fig3_10_001.dat};
        \addlegendentry{$\alpha=0.01$}

        \addplot[
            black,
            no marks,
            line width = 1pt,
            solid
        ]
        table {figures_tr/Fig3_10_005.dat};
        \addlegendentry{$\alpha=0.05$}

        \addplot[
            blue,
            no marks,
            line width = 1pt,
            solid
        ]
        table {figures_tr/Fig3_10_01.dat};
        \addlegendentry{$\alpha=0.1$}

        \addplot[
            red,
            only marks,
            mark=triangle*,
            mark size=3,
            mark repeat=1
        ]
        table {figures_tr/Fig3_10_001.dat};

        \addplot[
            black,
            only marks,
            mark=triangle*,
            mark size=3,
            mark repeat=1
        ]
        table {figures_tr/Fig3_10_005.dat};

        \addplot[
            blue,
            only marks,
            mark=triangle*,
            mark size=3,
            mark repeat=1
        ]
        table {figures_tr/Fig3_10_01.dat};

        \end{axis}
        \end{tikzpicture}
        \caption{}
    \end{subfigure}

    \begin{subfigure}{\linewidth}
        \centering
        \begin{tikzpicture}
        \begin{axis}[
            width=0.99\linewidth,
            xlabel = {$M$},
            ylabel = {$\overline{\eta}_r$},
            xmin = 1, xmax = 10,
            ymin = 0.6, ymax = 1,
            xtick = {1,2,3,4,5,6,7,8,9,10},
            ytick = {0.6,0.7,0.8,0.9,1},
            grid = major,
            legend image post style={xscale=0.9, every mark/.append style={solid}},
            legend cell align = {left},
            legend style={
                at={(1,0)},
                anchor=south east,
                font = \tiny
            }
        ]

        \addplot[
            red,
            no marks,
            line width = 1pt,
            solid
        ]
        table {figures_tr/Fig3_30_001.dat};
        \addlegendentry{$\alpha=0.01$}

        \addplot[
            black,
            no marks,
            line width = 1pt,
            solid
        ]
        table {figures_tr/Fig3_30_005.dat};
        \addlegendentry{$\alpha=0.05$}

        \addplot[
            blue,
            no marks,
            line width = 1pt,
            solid
        ]
        table {figures_tr/Fig3_30_01.dat};
        \addlegendentry{$\alpha=0.1$}

        \addplot[
            red,
            only marks,
            mark=triangle*,
            mark size=3,
            mark repeat=1
        ]
        table {figures_tr/Fig3_30_001.dat};

        \addplot[
            black,
            only marks,
            mark=triangle*,
            mark size=3,
            mark repeat=1
        ]
        table {figures_tr/Fig3_30_005.dat};

        \addplot[
            blue,
            only marks,
            mark=triangle*,
            mark size=3,
            mark repeat=1
        ]
        table {figures_tr/Fig3_30_01.dat};

        \end{axis}
        \end{tikzpicture}
        \caption{}
    \end{subfigure}

    \caption{PDE versus $M$ for $\gamma_t=90$ dB for different $\alpha$ values and a) $D_x=10$ m and b) $D_x=30$ m.}
    \label{Fig6}
\end{figure}

Finally, Figs.~\ref{Fig6}a and \ref{Fig6}b depict the PDE as a function of the number of PAs $M$ for different values of the waveguide attenuation coefficient $\alpha$, for deployment widths of $D_x = 10$~m and $D_x = 30$~m, respectively. As can be observed, increasing $\alpha$ leads to a slight degradation in the achievable PDE, reflecting the higher attenuation experienced along the waveguide, an effect that becomes more noticeable for larger deployment widths. However, for practical values of $\alpha$, the overall convergence behavior of the PDE with respect to $M$ remains largely unchanged, and the number of PAs required to achieve reliable performance does not vary significantly. In particular, the PDE exhibits a similar saturation trend across all considered $\alpha$ values, indicating that while waveguide attenuation affects the absolute performance level, it does not substantially impact the PA design in terms of the required number of PAs.

\section{Conclusion}
In this work, a comprehensive analytical framework was developed to characterize the performance of two-state PASs through the derivation of closed-form expressions for the outage probability and the ergodic achievable data rate. Specifically, the proposed analysis explicitly accounts for the discrete nature of the PA locations along the waveguide, enabling a direct comparison with the ideal continuous PAS benchmark. The obtained results demonstrated that for compact deployments, the ergodic data rate and the PDE quickly saturate with the number of PAs, indicating that only a limited number of PAs is sufficient to achieve near-continuous performance. In contrast, for larger deployment widths, additional PAs become increasingly beneficial, as they more effectively compensate for the increased propagation distances, although the required number of PAs remains moderate. Finally, the impact of waveguide attenuation was shown to affect the absolute performance levels without significantly altering the convergence behavior or the PA design requirements, highlighting the robustness of the two-state PAS architecture under practical conditions. Overall, the presented results confirmed that near-continuous PAS performance can be achieved with a finite and practical number of PAs, providing valuable theoretical insights and quantitative design guidelines for the efficient deployment of PASs in PWEs.

\appendices
\section{Calculation of Probability $P_l(\Delta)$}\label{App:1}
Below, we provide the derivation of the closed-form expressions of $P_l(\Delta)$ reported in Table~\ref{TableOut}. 
\subsubsection{$A_{0,k} \leq \min\!\left(\Delta^{2},\tfrac{D_y^{2}}{4}\right)$}
In this regime, $\sqrt{A_{0,k}} \leq \Delta$ and $A_{0,k}-\varepsilon^{2} \leq \tfrac{D_y^{2}}{4}$ for all $\varepsilon \in [0,\Delta]$, hence $P_l(\Delta)$ can be expressed as
\begin{equation}
\small
\label{Pl_general_case1}
\begin{aligned}
P_l(\Delta)
&=
\frac{1}{\Delta D_y}
\int_{0}^{\sqrt{A_{0,k}}}
\!\left[\!
\int_{-\frac{D_y}{2}}^{-\sqrt{A_{0,k}-\varepsilon^{2}}} \!\!\!\!\!\!\!dy_m
\!+
\int_{\sqrt{A_{0,k}-\varepsilon^{2}}}^{\frac{D_y}{2}} dy_m
\right]\!
d\varepsilon
\\[1mm]
&\quad+
\frac{1}{\Delta D_y}
\int_{\sqrt{A_{0,k}}}^{\Delta}
\int_{-\frac{D_y}{2}}^{\frac{D_y}{2}}
dy_m \, d\varepsilon .
\end{aligned}
\end{equation}
where, after some algebraic manipulations, \eqref{Pl_general_case1} can be written as
\begin{equation}
\small
\label{Pl_simplified_1}
    P_l(\Delta)
    =
    1 - \frac{2}{\Delta D_y}
    \int_{0}^{\sqrt{A_{0,k}}}
    \sqrt{A_{0,k}-\varepsilon^{2}} \,
    d\varepsilon.
\end{equation}
By setting $\theta=\arcsin\left( \frac{\varepsilon}{\sqrt{A_{0,k}}}\right)$, which implies 
$d\varepsilon = \sqrt{A_{0,k}}\cos\theta\,d\theta$ and 
$\sqrt{A_{0,k}-\varepsilon^{2}} = \sqrt{A_{0,k}}\cos\theta$, 
\eqref{Pl_simplified_1} can be rewritten as
\begin{equation}
\small
\label{Pl_simplified_2}
    P_l(\Delta)
    =
    1-\frac{2}{\Delta D_y}
    \int_{0}^{\frac{\pi}{2}}
    A_{0,k}\cos^{2}\!\theta \,
    d\theta,
\end{equation}
which yields
\begin{equation}
\small
    P_l(\Delta)
    =
    1-\frac{\pi A_{0,k}}{2 D_y \Delta},
\end{equation}
and coincides with the second row of Table~\ref{TableOut}.
\vspace{+2mm}
\subsubsection{$\Delta^{2}\leq A_{0,k} \leq \tfrac{D_y^{2}}{4}$}
In this case, $\sqrt{A_{0,k}}$ lies outside the interval $[0,\Delta]$, which implies that $A_{0,k}-\varepsilon^{2} \geq 0$ for all $\varepsilon \in [0,\Delta]$. Consequently, the second integral term in \eqref{eq:Pl_general_integral} vanishes, and 
$P_l(\Delta)$ reduces to
\begin{equation}
\label{Pl_general_case1}
\begin{aligned}
P_l(\Delta)
&=
\frac{1}{\Delta D_y}
\int_{0}^{\Delta}
\!\left[\!
\int_{-\frac{D_y}{2}}^{-\sqrt{A_{0,k}-\varepsilon^{2}}} \!\!\!\!\!\!\!dy_m
\!+
\int_{\sqrt{A_{0,k}-\varepsilon^{2}}}^{\frac{D_y}{2}} dy_m
\right]\!
d\varepsilon ,
\end{aligned}
\end{equation}
where after algebraic manipulations, it can be rewritten as
\begin{equation}
\small
\label{Pl_case2_start}
P_l(\Delta)
=
\frac{1}{\Delta D_y}
\int_{0}^{\Delta}
\left( D_y - 2\sqrt{A_{0,k}-\varepsilon^{2}} \right)
d\varepsilon.
\end{equation}
After rearranging terms, \eqref{Pl_case2_start} becomes
\begin{equation}
\label{Pl_case2_split}
P_l(\Delta)
=
1 - \frac{2}{\Delta D_y}
\int_{0}^{\Delta}
\sqrt{A_{0,k}-\varepsilon^{2}}\, d\varepsilon.
\end{equation}
Similarly with the previous case, by applying the substitution
$\theta=\arcsin\left( \frac{\varepsilon}{\sqrt{A_{0,k}}}\right)$, \eqref{Pl_case2_split} can be rewritten as
\begin{equation}
\label{Pl_case2_cos2}
P_l(\Delta)= 1 - \frac{2A_{0,k}}{\Delta D_y} \int_{0}^{\theta_{2}}
\cos^{2}\theta\, d\theta.
\end{equation}
where $\theta_{2}=\arcsin\left(\frac{\Delta}{\sqrt{A_{0,k}}}\right)$. Moreover, by using the trigonometric identity $\cos^{2}\theta = \tfrac{1 + \cos(2\theta)}{2}$ and after some algebraic manipulations, \eqref{Pl_case2_cos2} becomes
\begin{equation}
\small
\label{Pl_case2_integrated}
\begin{aligned}
P_l(\Delta)
= 1
&-\frac{A_{0,k}}{\Delta D_y}
\arcsin\!\left(\frac{\Delta}{\sqrt{A_{0,k}}}\right)  \\
&-\frac{A_{0,k}}{2 \Delta D_y}
\sin\!\left(2\arcsin\!\left(\frac{\Delta}{\sqrt{A_{0,k}}}\right)\right).
\end{aligned}
\end{equation}
Furthermore, by utilizing the identities $\sin(2\theta) =2\sin(\theta)\cos(\theta)$ and $\cos\left(\arcsin\left(x\right)\right)=\sqrt{1-x^2}$, after some algebraic manipulations, we obtain 
\begin{equation}
\small
P_l(\Delta)
=
1
-
\frac{A_{0,k}}{D_y \Delta}
\arcsin\!\left(\frac{\Delta}{\sqrt{A_{0,k}}}\right)
-
\frac{\sqrt{A_{0,k}-\Delta^{2}}}{D_y},
\end{equation}
which corresponds to the third row of Table~\ref{TableOut}.
\vspace{+2mm}
\subsubsection{$\tfrac{D_y^{2}}{4}\leq A_{0,k} \leq \Delta^{2}$}
In this regime, the equation $A_{0,k}-\varepsilon^{2} = \tfrac{D_y^{2}}{4}$ admits a solution $\varepsilon = \sqrt{A_{0,k}-\tfrac{D_y^{2}}{4}}$ in $[0,\Delta]$. For $\varepsilon \in \big[0,\sqrt{A_{0,k}-\tfrac{D_y^{2}}{4}}\big)$ the quantity $A_{0,k}-\varepsilon^{2}$ exceeds $\tfrac{D_y^{2}}{4}$, hence the condition $y_m^{2} \geq A_{0,k}-\varepsilon^{2}$ cannot be satisfied within $\left[-\tfrac{D_y}{2},\tfrac{D_y}{2}\right]$ and the contribution to $P_l(\Delta)$ is zero. Additionally, for $\varepsilon \in \big[\sqrt{A_{0,k}-\tfrac{D_y^{2}}{4}},\sqrt{A_{0,k}}\big]$ the feasible 
values of $y_m$ lie in $\big[-\tfrac{D_y}{2}, -\sqrt{A_{0,k}-\varepsilon^{2}}\big] \cup \big[\sqrt{A_{0,k}-\varepsilon^{2}}, \tfrac{D_y}{2}\big]$, while for $\varepsilon \in [\sqrt{A_{0,k}},\Delta]$ the whole interval $\left[-\tfrac{D_y}{2},\tfrac{D_y}{2}\right]$ is feasible. Therefore, $P_l(\Delta)$ can be written as
\begin{equation}
\label{Pl_case3_start}
\begin{aligned}
&P_l(\Delta)
=\frac{1}{\Delta D_y}
\int_{\sqrt{A_{0,k}-\tfrac{D_y^{2}}{4}}}^{\sqrt{A_{0,k}}}
\left( D_y - 2\sqrt{A_{0,k}-\varepsilon^{2}} \right)
d\varepsilon  \\
&+\frac{1}{\Delta}\int_{\sqrt{A_{0,k}}}^{\Delta} d\varepsilon,
\end{aligned}
\end{equation}
which can be equivalently written as
\begin{equation}
\label{Pl_case3_split}
P_l(\Delta)
=
1
-
\frac{2}{\Delta D_y}
\int_{\sqrt{A_{0,k}-\tfrac{D_y^{2}}{4}}}^{\sqrt{A_{0,k}}}
\sqrt{A_{0,k}-\varepsilon^{2}}\, d\varepsilon.
\end{equation}
Similarly to the previous case, by applying the substitution 
$\theta = \arcsin\!\left(\tfrac{\varepsilon}{\sqrt{A_{0,k}}}\right)$, \eqref{Pl_case3_split} can be rewritten as
\begin{equation}
\small
\label{Pl_case3_cos2}
P_l(\Delta)
=
1
-
\frac{2A_{0,k}}{\Delta D_y}
\int_{\theta_{3}}^{\frac{\pi}{2}}
\cos^{2}\theta\, d\theta,
\end{equation}
where $\theta_3=\arcsin\!\left(\frac{\sqrt{A_{0,k}-\tfrac{D_y^{2}}{4}}}{\sqrt{A_{0,k}}}\right)$.
By using the trigonometric identity $\cos^{2}\theta = \tfrac{1+\cos(2\theta)}{2}$ and following the same steps as in the second case, \eqref{Pl_case3_cos2} becomes
\begin{equation}
\begin{aligned}
P_l(\Delta)
&=
1-\frac{A_{0,k}}{\Delta D_y}\Bigg(\frac{\pi}{2}-\theta_3+\frac{1}{2}\sin(2\theta_3)\Bigg).
\end{aligned}
\end{equation}
Furthermore, by utilizing the identities $\sin(2\theta)=2\sin(\theta)\cos(\theta)$ and 
$\cos\big(\arcsin(x)\big)=\sqrt{1-x^{2}}$, after some algebraic manipulations we obtain
\begin{equation}
P_l(\Delta)
=
1
-
\frac{A_{0,k}}{D_y \Delta}
\arcsin\!\left(\frac{D_y}{2\sqrt{A_{0,k}}}\right)
-
\frac{\sqrt{A_{0,k}-\tfrac{D_y^{2}}{4}}}{2\Delta},
\end{equation}
which corresponds to the fourth row of Table~\ref{TableOut}.
\vspace{+2mm}
\subsubsection{$\max\!\left(\Delta^{2},\, \tfrac{D_y^{2}}{4}\right)\leq A_{0,k} \leq \Delta^{2} + \tfrac{D_y^{2}}{4}$}
In this regime, both equations $A_{0,k}-\varepsilon^{2} = \tfrac{D_y^{2}}{4}$ and $A_{0,k}-\varepsilon^{2} = 0$ admit solutions in $[0,\Delta]$. In more detail, for $\varepsilon \in \big[0,\sqrt{A_{0,k}-\tfrac{D_y^{2}}{4}}\big)$, the quantity $A_{0,k}-\varepsilon^{2}$ exceeds $\tfrac{D_y^{2}}{4}$ and the outage condition $y_m^{2} \geq A_{0,k}-\varepsilon^{2}$ cannot be satisfied within $\left[-\tfrac{D_y}{2},\tfrac{D_y}{2}\right]$, while for $\varepsilon \in \big[\sqrt{A_{0,k}-\tfrac{D_y^{2}}{4}},\Delta\big]$ we have $0 \leq A_{0,k}-\varepsilon^{2} \leq \tfrac{D_y^{2}}{4}$, so the feasible values of $y_m$ lie in $y_m \in \left[-\frac{D_y}{2}, -\sqrt{A_{0,k}-\varepsilon^{2}}\right]\cup\left[\sqrt{A_{0,k}-\varepsilon^{2}}, \frac{D_y}{2}\right]$, and there is no interval where $A_{0,k}-\varepsilon^{2}<0$. Therefore, \eqref{eq:Pl_general_integral} reduces to
\begin{equation}
\small
\label{Pl_case4_start}
P_l(\Delta)
=
\frac{1}{\Delta D_y}
\int_{\sqrt{A_{0,k}-\tfrac{D_y^{2}}{4}}}^{\Delta}
\left( D_y - 2\sqrt{A_{0,k}-\varepsilon^{2}} \right)
d\varepsilon.
\end{equation}
which can be also written as
\begin{equation}
\small
\label{Pl_case4_split}
P_l(\Delta)
=
1
-
\frac{2}{\Delta D_y}
\int_{\sqrt{A_{0,k}-\tfrac{D_y^{2}}{4}}}^{\Delta}
\sqrt{A_{0,k}-\varepsilon^{2}}\, d\varepsilon.
\end{equation}
Similarly to the previous cases, by applying the substitution $\theta = \arcsin\!\left(\frac{\varepsilon}{\sqrt{A_{0,k}}}\right)$, \eqref{Pl_case4_split} can be rewritten as
\begin{equation}
\label{Pl_case4_cos2}
P_l(\Delta)
=
1
-
\frac{2A_{0,k}}{\Delta D_y}
\int_{\theta_{3}}^{\theta_{2}}
\cos^{2}\theta\, d\theta,
\end{equation}
and by using the trigonometric identity 
$\cos^{2}\theta = \tfrac{1+\cos(2\theta)}{2}$ and following the same steps as in the previous cases, \eqref{Pl_case4_cos2} becomes
\begin{equation}
\small
\label{Pl_case4_integrated}
\begin{aligned}
P_l(\Delta)
&=
1
-
\frac{A_{0,k}}{\Delta D_y}
\Big(
\theta_{2}-\theta_{3}
+
\frac{1}{2}\big(\sin(2\theta_{2})-\sin(2\theta_{3})\big)
\Big).
\end{aligned}
\end{equation}
Finally, by utilizing the identities 
$\sin(2\theta)=2\sin(\theta)\cos(\theta)$ and 
$\cos\big(\arcsin(x)\big)=\sqrt{1-x^{2}}$, and after some algebraic manipulations, we obtain
\begin{equation}
\begin{aligned}
&P_l(\Delta)=1-\frac{A_{0,k}}{D_y \Delta}\Bigg[\arcsin\!\left(\frac{\Delta}{\sqrt{A_{0,k}}}\right) \\
&-
\arcsin\!\left(
\frac{\sqrt{4A_{0,k}-D_y^{2}}}{2\sqrt{A_{0,k}}}
\right)
\Bigg]
- \frac{1}{4\Delta}\sqrt{4A_{0,k}-D_y^{2}} \\&
- \frac{1}{D_y}\sqrt{A_{0,k}-\Delta^{2}}.
\end{aligned}
\end{equation}
which corresponds to the fifth row of Table~\ref{TableOut}.

\section{Calculation of Integrals $I_i$ and $I_j$}\label{App:A}
Below we provide the calculations for both $I_i$ and $I_j$ integrals:
\subsubsection{Integral $I_i$}
By setting $q = x+\tfrac{\delta^2}{4}$ and applying integration by parts with \(u=\ln(q+y_m^2)\) and \(dv=dy_m\) yields
\begin{equation}
I_i(x)=\Delta\Big[y_m\ln(q+y_m^2)\Big]_0^{D_y/2}
-2\Delta\int_{0}^{D_y/2}\frac{y_m^2}{q+y_m^2}\,dy_m,
\end{equation}
which can be rewritten as
\begin{equation}\label{I1_15}
\small
\begin{split}
    I_i(x)= & \Delta\Big[y_m\ln(q+y_m^2)\Big]_0^{D_y/2}
-2\Delta\int_{0}^{D_y/2}1\,dy_m  \\& + 2\Delta\int_{0}^{D_y/2}\frac{q}{q+y_m^2}\,dy_m.
\end{split}
\end{equation}
Moreover, by setting $u=\frac{y_m}{\sqrt{q}}$, \eqref{I1_15} can be written as
\begin{equation}\label{I1_16}
\small
\begin{split}
    I_i(x)= \frac{\Delta D_y}{2}
    \ln\!\left(\frac{D_y^2}{4}+q\right)
\!-\Delta D_y 
+ 2\Delta\sqrt{q}\!\int_{0}^{\frac{D_y}{2\sqrt{q}}}\!\frac{1}{1+u^2}\,du.
\end{split}
\end{equation}
Finally, by utilizing \cite[Eq.~(2.01/15)]{GradshteynRyzhik2014}, \eqref{Iix} can be derived, which completes the derivation of $I_i$.
\subsubsection{Integral $I_j$}
By setting $\tan^{-1} \! \left( \frac{\Delta}{\sqrt{x+ y^2_m}}\right)=u$ and $dv=2 \sqrt{x+ y^2_m}\,dy_m$ and applying integration by parts, \eqref{Ij_13} can be written as
\begin{equation}\label{Ij_17}
\small
\begin{aligned}
&I_j(x)= \!\int_{0}^{\frac{D_y}{2}}
\frac{\Delta\,y_m
\Big(y_m\sqrt{x+y_m^2}
+ x\ln\!\big(y_m+\sqrt{x+y_m^2}\big)\!\Big)}
{\sqrt{x+y_m^2}\,\big(x+y_m^2+\Delta^2\big)}\,dy_m \\
& \!+\!\Bigg[\!
\tan^{-1}\!\left(\!\tfrac{\Delta}{\sqrt{x+y_m^2}}\!\right)
\!\Big(y_m\!\sqrt{x+y_m^2}
\!+\! x\ln\!\Big(y_m+ \sqrt{x+y_m^2}\Big)\!\Big)  \!\Bigg]_{0}^{\!\frac{D_y}{2}}\!,
\end{aligned}
\end{equation}
which, after some algebraic manipulations, can be expressed as
\begin{equation}\label{Ij_18}
\small
\begin{aligned}
&I_j(x)\!= \!\tan^{-1}\!\left(\!\frac{\Delta}{\sqrt{x+\tfrac{D_y^2}{4}}}\!\right)
\!\Bigg(\!\frac{D_y}{2}\sqrt{x+\tfrac{D_y^2}{4}}
+ \\&x\ln\!\Big(\tfrac{D_y}{2}+\sqrt{x+\tfrac{D_y^2}{4}}\Big)\!\Bigg) 
- x\ln(\sqrt{x})\,\tan^{-1}\!\left(\frac{\Delta}{\sqrt{x}}\right)\!+ \\[2pt]
& \!\underbrace{\displaystyle\int\limits_{0}^{\frac{D_y}{2}}\!
\frac{\Delta\,y_m^2}{x+y_m^2+\Delta^2}dy_m}_{\textstyle J_1}
+
\underbrace{\displaystyle\int\limits_{0}^{\frac{D_y}{2}}
\frac{\Delta\,x\,y_m\,\ln\!\big(y_m+\sqrt{x+y_m^2}\big)}
{\sqrt{x+y_m^2}\,\big(x+y_m^2+\Delta^2\big)}dy_m}_{\textstyle J_2}\!.
\end{aligned}
\end{equation}
The evaluation of $I_j(x)$ then proceeds through the calculation of $J_1$ and $J_2$, each of which can be expressed in closed form.

\underline{\textit{i) Calculation of $J_1$:}} Initially, to calculate $J_1$ we can reformulate it as follows
\begin{equation}
\small
    J_1=\Delta\int_{0}^{\frac{D_y}{2}}\!1-
\frac{x+ \Delta^2}{x+y_m^2+\Delta^2}dy_m,
\end{equation}
Moreover, by setting $q = x+\Delta^2$ and $u = \tfrac{y_m}{\sqrt{q}}$, after some algebraic manipulations, we obtain
\begin{equation}
\small
    J_1=\Delta\!\left[\frac{D_y}{2}
    - \sqrt{q}\!\int_{0}^{\frac{D_y}{2\sqrt{q}}}\!\frac{1}{1+u^2}\,du\right].
\end{equation}
Finally, by using \cite[(2.01/15)]{GradshteynRyzhik2014}, yields
\begin{equation}\label{J1_final}
\small
    J_1 = \frac{\Delta D_y}{2}
    - \Delta\sqrt{x+\Delta^2}\,
    \tan^{-1}\!\left(\frac{D_y}{2\sqrt{x+\Delta^2}}\right),
\end{equation}
which completes the calculation of $J_1$.

\underline{\textit{ii) Calculation of $J_2$:}} 
By setting $y_m=\!\sqrt{x}\sinh(u)$, then $J_2$ can be expressed as
\begin{equation}\label{J2_22}
\small
J_2= \displaystyle\int\limits_{0}^{\operatorname{asinh}\!\left(\frac{D_y}{2\sqrt{x}}\right)} 
    \frac{\Delta x\sqrt{x}\sinh(u)\!\ln\!\Big(\!\sqrt{x}\sinh(u)\!+\!\sqrt{x}\cosh(u)\!\Big)}
    {\big(x\cosh^2(u)+\Delta^2\big)}du.
\end{equation}
Additionally, by utilizing the hyperbolic identity $\sinh(u)+\cosh(u)=e^{u}$, after some algebraic manipulations, \eqref{J2_22} can be rewritten as
\begin{equation}
\small
J_2= \Delta \sqrt{x}\displaystyle\int\limits_{0}^{\operatorname{asinh}\left(\frac{D_y}{2\sqrt{x}}\right)}
\frac{\sinh (u)\big(\ln(\sqrt{x})+u\big)}
{\cosh^2(u)+\frac{\Delta^2}{x}}\,du.
\end{equation}
By applying integration by parts with 
$s=\ln(\sqrt{x})+u$ and 
$dw=\Delta\sqrt{x}\frac{\sinh u}{\cosh^2 u+\frac{\Delta^2}{x}}\,du$, 
we obtain
\begin{equation}\label{J2_24}
\small
\begin{split}
J_2= &\left[x\Big(\ln(\sqrt{x})+u\Big)\tan^{-1}\!\Big(\tfrac{\sqrt{x}}{\Delta}\cosh(u)\Big)
\right]_{0}^{\operatorname{asinh}\!\left(\frac{D_y}{2\sqrt{x}}\right)} \\&
- x\!\int_{0}^{\operatorname{asinh}\!\left(\frac{D_y}{2\sqrt{x}}\right)}
\tan^{-1}\!\Big(\tfrac{\sqrt{x}}{\Delta}\cosh(u)\Big)\,du.
\end{split}
\end{equation}
By utilizing the identity $\cosh(\operatorname{asinh} z)=\sqrt{1+z^2}$, then \eqref{J2_24} simplifies to
\begin{equation}\label{J2_25}
\small
\begin{aligned}
J_2 & = x\Bigg(\!\Big(\!\ln(\sqrt{x})+\operatorname{asinh}\!\left(\tfrac{D_y}{2\sqrt{x}}\right)\!\Big)\!
\tan^{-1}\!\left(\!\tfrac{\sqrt{x+\frac{D_y^2}{4}}}{\Delta}\!\right) \!-\!\ln(\sqrt{x}) \\
& \times\tan^{-1}\!\Big(\tfrac{\sqrt{x}}{\Delta}\Big)
-\int_{0}^{\operatorname{asinh}\!\left(\frac{D_y}{2\sqrt{x}}\right)}
\tan^{-1}\!\Big(\tfrac{\sqrt{x}}{\Delta}\cosh(u)\Big)du \Bigg).
\end{aligned}
\end{equation}
By using the expansion $\tan^{-1}(z)=\tfrac{j}{2}\big[\ln(1-jz)-\ln(1+jz)\big]$, then \eqref{J2_25} can be reformulated as
\begin{equation}\label{J2_26}
\small
\begin{aligned}
J_2 &= x\Bigg(\!\Big(\!\ln(\sqrt{x})+\operatorname{asinh}\!\left(\tfrac{D_y}{2\sqrt{x}}\right)\!\Big)\!
\tan^{-1}\!\left(\!\tfrac{\sqrt{x+\frac{D_y^2}{4}}}{\Delta}\!\right) \!-\!\ln(\sqrt{x}) \\
& \times\tan^{-1}\!\Big(\tfrac{\sqrt{x}}{\Delta}\Big)
- \frac{j}{2}\!\displaystyle\int\limits_{0}^{\operatorname{asinh}\!\left(\tfrac{D_y}{2\sqrt{x}}\right)}\!
\ln\!\Bigg(\frac{1-j\tfrac{\sqrt{x}}{\Delta}\cosh(u)}{1+j\tfrac{\sqrt{x}}{\Delta}\cosh(u)}\Bigg)du
\Bigg).
\end{aligned}
\end{equation}
Additionally, by setting $\xi=e^{u}$ and after some algebraic manipulations, \eqref{J2_26} becomes
\begin{equation}\label{J2_27}
\small
\begin{aligned}
J_2 & = x\Bigg(\!\Big(\!\ln(\sqrt{x})+\operatorname{asinh}\!\left(\tfrac{D_y}{2\sqrt{x}}\right)\!\Big)\!
\tan^{-1}\!\left(\!\tfrac{\sqrt{x+\frac{D_y^2}{4}}}{\Delta}\!\right) \!-\!\ln(\sqrt{x}) \\
& \times\tan^{-1}\!\Big(\tfrac{\sqrt{x}}{\Delta}\Big)
- \frac{j}{2}\int\limits_{1}^{e^{A}}
\ln\!\Bigg(\frac{2\xi - j\,\tfrac{\sqrt{x}}{\Delta}\,(\xi^2+1)}
     {2\xi + j\,\tfrac{\sqrt{x}}{\Delta}\,(\xi^2+1)}\Bigg)\ \frac{d\xi}{\xi}\Bigg),
\end{aligned}
\end{equation}
with $A={\mathrm{asinh}}\left(\tfrac{D_y}{2\sqrt{x}}\right)$, and by factorizing the quadratic terms in the numerator and denominator of the logarithm in \eqref{J2_27}, the logarithm can be expressed as four elementary logarithms, resulting in
\begin{equation}\label{J2_28}
\small
\begin{aligned}
&J_2= x\Bigg(\!\Big(\!\ln(\sqrt{x})+\operatorname{asinh}\!\left(\tfrac{D_y}{2\sqrt{x}}\right)\!\Big)\!
\tan^{-1}\!\left(\!\tfrac{\sqrt{x+\frac{D_y^2}{4}}}{\Delta}\!\right) \!-\!\ln(\sqrt{x}) \\
&\!\times\tan^{-1}\!\Big(\tfrac{\sqrt{x}}{\Delta}\Big)
- \frac{j}{2}\int\limits_{1}^{e^A}\!\
\ln\!\Big(\xi + j\,\tfrac{\Delta}{\sqrt{x}}\big(1+\sqrt{1+\tfrac{x}{\Delta^2}}\big)\Big) \\
&\!+\!\ln\!\Big(\xi - j\,\tfrac{\Delta}{\sqrt{x}}\big(\sqrt{1+\tfrac{x}{\Delta^2}}-1\big)\Big)\!-\!\ln\!\Big(\xi - j\,\tfrac{\Delta}{\sqrt{x}}\big(1-\sqrt{1+\tfrac{x}{\Delta^2}}\big)\Big) \\
&\!-\ln\!\Big(\xi - j\,\tfrac{\Delta}{\sqrt{x}}\big(1+\sqrt{1+\tfrac{x}{\Delta^2}}\big)\Big)
\Big]\frac{d\xi}{\xi}\Bigg).
\end{aligned}
\end{equation}
Moreover, by applying the identity $\ln(\xi - a)=\ln\xi+\ln\!\big(1-\tfrac{a}{\xi}\big)$, \eqref{J2_28} can be rewritten as
\begin{equation}\label{J2_29}
\small
\begin{aligned}
&J_2= x\Bigg(\!\Big(\!\ln(\sqrt{x})+\operatorname{asinh}\!\left(\tfrac{D_y}{2\sqrt{x}}\right)\!\Big)\!
\tan^{-1}\!\left(\!\tfrac{\sqrt{x+\frac{D_y^2}{4}}}{\Delta}\!\right) \!-\!\ln(\sqrt{x}) \\
&\times\tan^{-1}\!\Big(\tfrac{\sqrt{x}}{\Delta}\Big)
- \frac{j}{2}\!\int\limits_{1}^{e^A}\!
\Big[
\ln\!\Big(1-\tfrac{-\,j\,\tfrac{\Delta}{\sqrt{x}}\big(1+\sqrt{1+\tfrac{x}{\Delta^2}}\big)}{\xi}\Big) \\
&+\ln\!\Big(1-\tfrac{+\,j\,\tfrac{\Delta}{\sqrt{x}}\big(\sqrt{1+\tfrac{x}{\Delta^2}}-1\big)}{\xi}\Big)
-\ln\!\Big(1-\tfrac{+\,j\,\tfrac{\Delta}{\sqrt{x}}\big(1-\sqrt{1+\tfrac{x}{\Delta^2}}\big)}{\xi}\Big) \\
&\!-\ln\!\Big(1-\tfrac{+\,j\,\tfrac{\Delta}{\sqrt{x}}\big(1+\sqrt{1+\tfrac{x}{\Delta^2}}\big)}{\xi}\Big)
\Big]\frac{d\xi}{\xi}\Bigg).
\end{aligned}
\end{equation}
By taking into account that $\int \frac{\ln\!\big(1-\tfrac{a}{\xi}\big)}{\xi}\,d\xi=\operatorname{Li}_2\!\Big(\tfrac{a}{\xi}\Big)$, we obtain
\begin{equation}\label{J2_30}
\small
\begin{aligned}
&J_2= x\Bigg(\!\Big(\!\ln(\sqrt{x})+\operatorname{asinh}\!\left(\tfrac{D_y}{2\sqrt{x}}\right)\!\Big)\!
\tan^{-1}\!\left(\!\tfrac{\sqrt{x+\frac{D_y^2}{4}}}{\Delta}\!\right) \!-\!\ln(\sqrt{x}) \\
&\times\tan^{-1}\!\Big(\tfrac{\sqrt{x}}{\Delta}\Big)
- \frac{j}{2}\Bigg[
\operatorname{Li}_2\!\Big(\tfrac{-\,j\,\tfrac{\Delta}{\sqrt{x}}\big(1+\sqrt{1+\tfrac{x}{\Delta^2}}\big)}{\xi}\Big)
\\
&+\operatorname{Li}_2\!\Big(\tfrac{+\,j\,\tfrac{\Delta}{\sqrt{x}}\big(\sqrt{1+\tfrac{x}{\Delta^2}}-1\big)}{\xi}\Big)-\operatorname{Li}_2\!\Big(\tfrac{+\,j\,\tfrac{\Delta}{\sqrt{x}}\big(1-\sqrt{1+\tfrac{x}{\Delta^2}}\big)}{\xi}\Big) \\
&-\operatorname{Li}_2\!\Big(\tfrac{+\,j\,\tfrac{\Delta}{\sqrt{x}}\big(1+\sqrt{1+\tfrac{x}{\Delta^2}}\big)}{\xi}\Big)
\Bigg]_{1}^{e^{A}}\Bigg),
\end{aligned}
\end{equation}
and by using the identities $\operatorname{Li}_2\!\Big(\tfrac{1}{z}\Big)
= -\operatorname{Li}_2(z)-\frac{\pi^2}{6}-\frac{1}{2}\,\ln^2(-z)$, and $\ln(-z)=\ln z + j\pi$, after some algebraic manipulations we obtain
\begin{equation}\label{J2_31}
\small
\begin{aligned}
&J_2= x\Bigg(\!\Big(\!\ln(\sqrt{x})+\operatorname{asinh}\!\left(\tfrac{D_y}{2\sqrt{x}}\right)\!\Big)\!
\tan^{-1}\!\left(\!\tfrac{\sqrt{x+\frac{D_y^2}{4}}}{\Delta}\!\right) \!-\!\ln(\sqrt{x}) \\
&\!\times\tan^{-1}\!\Big(\tfrac{\sqrt{x}}{\Delta}\Big)
-\frac{\pi}{2}\operatorname{asinh}\!\left(\tfrac{D_y}{2\sqrt{x}}\right) \\
&\!-\frac{j}{2}\Bigg[
\operatorname{Li}_2\!\Big(\tfrac{-j\Delta}{\sqrt{x}}\!\left(1+\sqrt{1+\tfrac{x}{\Delta^2}}\right)e^{-\!A}\Big)
\\&\!-\operatorname{Li}_2\!\Big(\tfrac{-j\Delta}{\sqrt{x}}\!\left(1+\sqrt{1+\tfrac{x}{\Delta^2}}\right)e^{A}\Big)
-\operatorname{Li}_2\!\Big(\tfrac{j\Delta}{\sqrt{x}}\!\left(1+\sqrt{1+\tfrac{x}{\Delta^2}}\right)e^{-\!A}\Big)
\\
&\!+\operatorname{Li}_2\!\Big(\tfrac{j\Delta}{\sqrt{x}}\!\left(1+\sqrt{1+\tfrac{x}{\Delta^2}}\right)e^{A}\Big)
\Bigg]\Bigg).
\end{aligned}
\end{equation}
Finally, by utilizing the definition of the arctangent integral function $\mathrm{Ti}_2(z)$, we obtain \eqref{J2_32}, which concludes the derivation of $J_2$.
\begin{figure*}[t]
\centering
\begin{equation}\label{J2_32}
\small
\begin{aligned}
&J_2= x\!\Bigg[\!\Big(\!\ln(\sqrt{x})+\operatorname{asinh}\!\left(\tfrac{D_y}{2\sqrt{x}}\right)\!\Big)\!
\tan^{-1}\!\left(\!\tfrac{\sqrt{x+\frac{D_y^2}{4}}}{\Delta}\!\right) \!-\!\ln(\sqrt{x})\tan^{-1}\!\Big(\tfrac{\sqrt{x}}{\Delta}\Big)
-\tfrac{\pi}{2}\operatorname{\mathrm{asinh}}\!\left(\tfrac{D_y}{2\sqrt{x}}\right) 
-\mathrm{Ti}_2\Big(
e^{\operatorname{\mathrm{asinh}}\!\left(\tfrac{D_y}{2\sqrt{x}}\right)}
\tfrac{\Delta}{\sqrt{x}}
\!\left(\sqrt{1+\tfrac{x}{\Delta^2}}-1\right)\!\Big) \\&
+\mathrm{Ti}_2\Big(
\tfrac{\Delta}{\sqrt{x}}
\!\left(\sqrt{1+\tfrac{x}{\Delta^2}}-1\right)\!\Big)
-\mathrm{Ti}_2\Big(
-\,e^{\operatorname{asinh}\!\left(\tfrac{D_y}{2\sqrt{x}}\right)}
\tfrac{\Delta}{\sqrt{x}}
\!\left(\sqrt{1+\tfrac{x}{\Delta^2}}+1\right)\!\Big)+\mathrm{Ti}_2\Big(
-\,\tfrac{\Delta}{\sqrt{x}}
\!\left(\sqrt{1+\tfrac{x}{\Delta^2}}+1\right)\!\Big)
\Bigg],
\end{aligned}
\end{equation}
\vspace{2mm}
\hrule
\end{figure*}

\bibliographystyle{IEEEtran}
\bibliography{Bibliography}
\end{document}